\documentclass[aps,prl,twocolumn,superscriptaddress,amsmath,amssymb,groupedaddress,bibnotes,nobalancelastpage]{revtex4-2}
\usepackage[english]{babel}
\usepackage{graphicx,bbm}
\usepackage[utf8]{inputenc}
\usepackage[colorlinks=true,linkcolor=blue, citecolor=blue, urlcolor=blue, bookmarks]{hyperref}
\usepackage{bbold}
\usepackage{physics}
\usepackage[normalem]{ulem}

\usepackage[dvipsnames]{xcolor}
\usepackage[scr=boondox]{mathalfa}
\newcommand{\bfp}{\mathbf{p}}
\newcommand{\bfh}{\mathbf{h}}
\newcommand{\bfl}{\boldsymbol{\lambda}}

\newcommand{\Partial}[3]{\left( \frac{\partial #1}{\partial #2} \right)_{#3}}

\begin{document} 
\title{Navier-Stokes equations for nearly integrable quantum gases}

\author{Maciej {\L}ebek}
\email{maciej.lebek@fuw.edu.pl}
\author{Miłosz Panfil}
\email{milosz.panfil@fuw.edu.pl}
\affiliation{Faculty of Physics, University of Warsaw, Pasteura 5, 02-093 Warsaw, Poland}

\date{\today}
%------------------------------------------------------------------------------------
\begin{abstract}
The Navier-Stokes equations are paradigmatic equations describing hydrodynamics of an interacting system with microscopic interactions encoded in transport coefficients. In this work we show how the Navier-Stokes equations arise from the microscopic dynamics of nearly integrable $1d$ quantum many-body systems. We build upon the recently developed hydrodynamics of integrable models to study the effective Boltzmann equation with collision integral taking into account the non-integrable interactions. We compute the transport coefficients and find that the resulting Navier-Stokes equations have two regimes, which differ in the viscous properties of the fluid. We illustrate the method by computing the transport coefficients for an experimentally relevant case of coupled $1d$ cold-atomic gases.
\end{abstract}
%------------------------------------------------------------------------------------
\maketitle

%==============================================================================
Hydrodynamics is a universal theory describing phenomena of transport and applicable to systems of sizes ranging from nuclear through biological up to cosmological scales \cite{Dorfman2021,relativistic_hydro,vogel1996life}. 
When viewed as an emergent theory, it assumes a huge reduction of the degrees of freedom present at the microscopic scale. The high-energy degrees of freedom are effectively integrated out which leads to dissipation effects present in the Navier-Stokes (NS) equations and captured by the transport coefficients. These are non-universal quantities depending on the details of a microscopic theory. In the present work we show how the NS equations arise from quantum-many body dynamics for systems that are nearly integrable. We also derive expressions for the transport coefficients which treat the integrable part of the interactions in an exact and non-perturbative way. 

Recent years have witnessed important developments in the field of non-equilibrium dynamics of isolated quantum-many body systems. 
The program was especially intriguing in 1+1 dimensions where interacting integrable models showcase behaviors differing from those of generic systems. Nonetheless, the rich methods of quantum integrability allowed for powerful descriptions~\cite{rigol_review,Essler_2016,QA_Caux,Mori_2018,Borsi_2021,Bouchoule2022}.
Importantly, these developments were fuelled by experimental progress in creating and manipulating cold-atomic systems~\cite{kinoshita2006quantum,2019PhRvL.122i0601S, Tang2018, Cataldini2022,Moller2021,Li2020,Li2023,Wilson2020}. It resulted in an understanding that typically quantum integrable systems relax to a generalized Gibbs ensemble (GGE)~\cite{rigol2008thermalization,gge_prl,langen2015experimental} which includes, beside the energy, also the additional conserved charges present in such theories. 

The presence of additional conservation laws affects also the hydrodynamics of integrable models. The resulting generalized hydrodynamics (GHD)~\cite{fagotti_ghd,doyon_ghd,JSTAT_GHD_review, Doyon2023,lecture_notes_GHD} takes a form of a coupled continuity equations for an extensive number of conserved charge densities. The dispersive processes are also taken into account as a diffusion intrinsic to the integrable dynamics~\cite{hydro_diff_prl,PhysRevB.98.220303,DeNardis2019}. As the hydrodynamic picture assumes that locally the system is in equilibrium, it requires a separation of time-scales between small $t_{\rm GGE}$ of the local thermalization to the GGE and large $t_{\rm GHD}$ associated with the flow created by spatial inhomogeneities.

\begin{figure}
    \centering
    \includegraphics[scale=0.48]{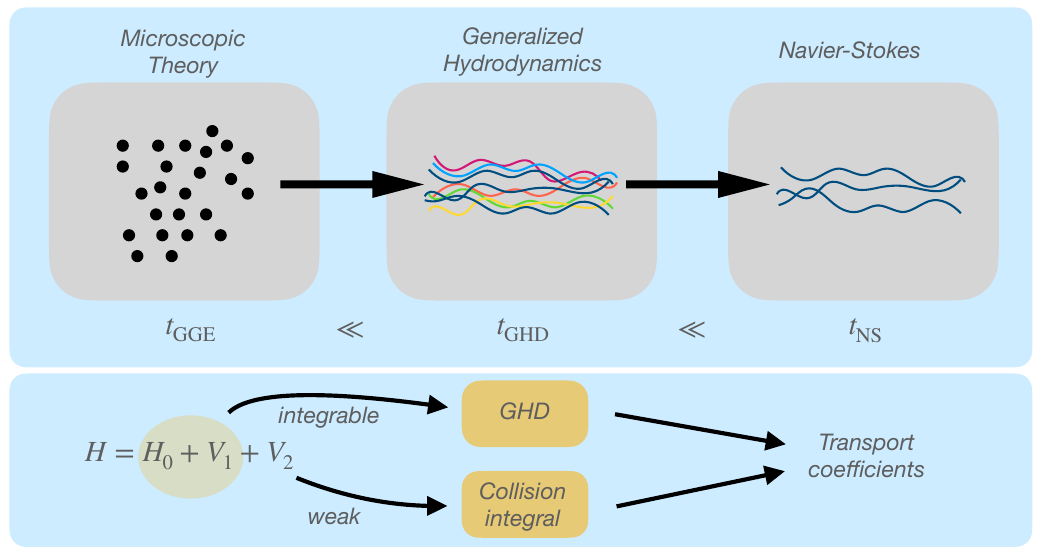}
    \caption{{\em top panel:} Typically quantum integrable systems locally relax to the GGE with some characteristic time-scale $t_{\rm GGE}$. When the system is globally inhomogeneous further evolution is described by the GHD and occurs at time-scale associated with dynamics of spatial gradients of conserved densities. Finally, in the presence of integrability breaking terms and at larger time-scales, the energy dissipates from higher conserved charges to the $3$ elementary ones resulting in the NS equations. {\em bottom panel:} Our method separates the full Hamiltonian into integrable Hamiltonian and weak integrability breaking term $V_2$. The transport coefficients have then two contributions, one from the interactions present in the integrable theory (integrable collisions) and second from the integrability breaking term (non-integrable collisions). }
    \label{fig:intro}
\end{figure}

This formalism can be extended beyond purely integrable models by capturing the additional interactions through the Boltzmann collision integral~\cite{weak_integ_breaking,Moller2021,PGK, Cataldini2022, Moller2021,Lopez2021,Lopez2022,Lopez2023,Caux2019, Bouchoule2020,Bastianello2020}. Building on that results one finds that there is an additional time scale, that we call $t_{\rm NS}$, associated with the integrability breaking terms. In this work we show that for models with Galilean invariance and with the new interaction term which conserves only the particle number, momentum and energy this results in the NS equations.  We also find that there are two contributions to the transport coefficients. They origin from two types of collisions present in the system. First are integrable collisions captured by the diffusion term of the GHD. Second are the integrability breaking collisions captured by the Boltzmann collision integral. This is summarized in Fig.~\ref{fig:intro}. Our approach to the problem generalizes the classic kinetic theory derivations of hydrodynamics through Chapman-Enskog expansion~\cite{ResiboisBOOK}.

%==============================================================================
{\bf Generalized hydrodynamics:} 
The GHD is a theory of hydrodynamics for integrable models, such as integrable spin chains or 1d Bose gas, where an extensive number of local conservation laws are present. Whereas the theory is fairly new~\cite{fagotti_ghd,doyon_ghd} (with earlier developments in~\cite{PhysRevLett.113.187203}), it has been studied quite extensively (e.g. see the collections of reviews~\cite{JSTAT_GHD_review}) and was verified numerically \cite{Doyon2017,Doyon2017a,hydro_diff_prl,Biagetti2024} and experimentally~\cite{2019PhRvL.122i0601S,malvania2021generalized,Moller2021}. Recently it was also shown how to extend its applicability, by the logic of {\em Boltzmann collision integral}, to the cases when the integrability is weakly broken~\cite{PhysRevB.101.180302,weak_integ_breaking,Bastianello_2021,Moller2021,PGK,prethermal_ladder,Lebek2024,Bertini2015,Bertini2016,DelVecchio2022}. 

The GHD assumes a local relaxation of the system to the GGE~\cite{Mossel2012,2012PhRvL.109q5301C} which in its microcanonical picture is described by the rapidity distribution $\rho_{\rm p} \equiv \rho_{\rm p}(\lambda; x,t)$ slowly varying in space and time. The rapidity distribution evolves through ballistic propagation of particles accompanied by integrable scattering processes leading to diffusive spreading and non-integrable scatterings captured by the collision integral, as described by the GHD equation
\begin{equation} \label{GHD_full}
	\partial_t \rho_{\rm p} + \partial_x \left(v_{\rho} \rho_{\rm p} \right) +\mathfrak{f} \partial_\lambda \rho_{\rm p}= \frac{1}{2} \partial_x \left(\mathfrak{D}_{\rho}\partial_x \rho_{\rm p} \right) + \mathcal{I}[\rho_{\rm p}],
\end{equation}
with effective velocity $v_{\rho}$, the diffusion kernel $\mathfrak{D}_{\rho}$ and the Boltzmann collision integral $\mathcal{I}[\rho_{\rm p}]$. The effective velocity fulfills the integral equation~\cite{doyon_ghd,lecture_notes_GHD} $v_{\rho}(\lambda)=\lambda+2 \pi \int {\rm d} \lambda' \mathcal{T}(\lambda-\lambda')\rho_{\rm p}(\lambda')(v_{\rho}(\lambda')-v_{\rho}(\lambda))$, where $\mathcal{T}(\lambda)$ is model-dependent scattering kernel. The diffusion kernel~\cite{hydro_diff_prl,DeNardis2019} can be found in the Appendix A. We also consider external potential $U(x)$ and associated force term $\mathfrak{f}= - \partial_xU$~\cite{Doyon2017force,Bastianello2020a,Durnin2021,Cao2018}.  The index $\rho$ signifies that the effective velocity and diffusion kernel are functionals of $\rho_{\rm p}$ and are space-time dependent. This renders the GHD equations highly nonlinear. Furthermore, the Boltzmann integral can be evaluated from the matrix elements of the perturbing operator in the spirit of the Fermi's golden rule~\cite{weak_integ_breaking, PGK}. However, for the large scale dynamics the crucial aspect are their collision invariants.

In what follows we quite generally assume that there are $3$ collision invariants corresponding to the total particle number, total momentum and energy such that 
\begin{equation} \label{coll_inv}
	\int {\rm d}\lambda\, \mathscr{h}_j(\lambda)\, \mathcal{I}[\rho_{\rm p}](\lambda) = 0, \quad j=0, 1, 2.
\end{equation}  
where $\mathscr{h}_j(\lambda)$ are single-particle eigenvalues of the conserved charges of the underlying integrable model,
\begin{equation} \label{charges_eiegenvalues}
	 \mathscr{q}_j = \langle \mathscr{h}_j \rangle \equiv \int {\rm d}\lambda\, \mathscr{h}_j(\lambda) \rho_{\rm p}(\lambda).
\end{equation} 
Functions $\mathscr{h}_j(\lambda)$ for a Galilean invariant theory are $\mathscr{h}_j(\lambda) = \lambda^j/j!$. The $3$ collisions invariants determine the nature of the stationary state of~\eqref{GHD_full} to be a homogeneous thermal (boosted) state and the corresponding $\rho_{\rm p}(\lambda)$ follows from the standard construction of the Thermodynamic Bethe Ansatz~\cite{Yang1969,Takahashi1999}, which we recall in the Appendix B.

%==============================================================================
{\bf Macroscopic conservation laws:}
The $3$ collision invariants lead to macroscopic conservation laws. These are conveniently formulated using hydrodynamic fields 
\begin{equation}
   u(x,t) = \frac{\mathscr{q}_1(x,t)}{\mathscr{q}_0 (x,t)},\quad   e(x,t) = \frac{\mathscr{q}_2(x,t)}{\mathscr{q}_0(x,t)} - \frac{1}{2} u(x,t)^2,
\end{equation}
and we additionally rename the particle
density field $\varrho(x,t) = \mathscr{q}_0(x,t)$. We will refer to the set of hydrodynamic fields as $\{\varrho_\alpha\} = {\varrho, u, e}$.  To this end, we multiply the GHD eq.~\eqref{GHD_full} by $\mathscr{h}_j(\lambda)$, integrate over $ \lambda$ and use the property~\eqref{coll_inv}. Following this procedure for $j=0,1,2$ we obtain 
\begin{equation} \label{NS_eqns}
	\begin{aligned}
		\partial_t \varrho &= - \partial_x (\varrho u), \qquad \partial_t (\varrho u) = - \partial_x (\varrho u^2 + \mathcal{P})+\mathfrak{f}\varrho, \\
		\partial_t (\varrho e) &= - \partial_x (u \varrho e+\mathcal{J}) - \mathcal{P} \partial_x u,
	\end{aligned}
\end{equation}
with $\mathcal{P}(x,t)$ local pressure and $\mathcal{J}(x,t)$ local heat current.
The pressure and the heat current have two contributions. One intrinsic to the diffusion term present in the GHD and given by $\mathcal{P}_\mathfrak{D} = \mathscr{h}_1 \mathfrak{D}_{\rho}\partial_x \rho_{\rm p}/2$ and $\mathcal{J}_\mathfrak{D} = \mathscr{h}_2 \mathfrak{D}_{\rho}\partial_x \rho_{\rm p}/2$. The second one resembles the standard expressions from the kinetic theory,
\begin{align}
	\mathcal{P}_v = \langle (\lambda - u) v(\lambda)\rangle,\quad 
	\mathcal{J}_v = \frac{1}{2} \langle (\lambda - u)^2 (v(\lambda)-u) \rangle.
\end{align}
The derivation of the conservation laws together with the formulas for $\mathcal{P}$ and $\mathcal{J}$ is elementary and presented in~\cite{SM}. We only note that in deriving the first conservation law the identity $\langle v \rangle = \langle \mathscr{h}_1 \rangle$, a consequence of the Galiliean invariance, is operatorial. This property establishes the velocity field $u$ a relevant hydrodynamic degree of freedom. 

The equations~\eqref{NS_eqns} are not closed because the pressure and heat current are functions of the full distribution $\rho_{\rm p}$ rather than just its $3$ moments encoded through fields $\varrho$, $u$ and $e$. To circumvent this problem, in the phenomenological treatment one assumes the following expressions 
\begin{align} \label{pheno}
	\mathcal{P} = P - \zeta \partial_x u, \qquad \mathcal{J} = - \kappa \partial_x T,
\end{align}
in terms of the hydrodynamic fields. Here $P=P(\varrho, e)$ is the hydrostatic pressure, $\zeta$ is the bulk viscosity and $\kappa$ is the thermal conductivity, respectively. The temperature field and the energy field are related through the thermodynamic identity $T^{-1} = (\partial s/\partial e)_{\varrho}$, where $s$ is thermodynamic entropy per particle~\cite{Yang1969,Mossel2012,Takahashi1999}. This leaves the theory with two phenomenological constants $\zeta$ and $\kappa$ called {\em transport coefficients}. In this way we obtain the one-dimensional version of NS equations~\cite{Balescu1975,spohn2012large, Chakraborti2021}, for the fields $\varrho(x,t)$, $u(x,t)$ and $e(x,t)$. In the following we confirm this phenomenological picture and provide explicit formulas for the transport coefficients.

%==============================================================================
{\bf Chapman-Enskog method:} The standard procedure to justify the phenomenological picture and to turn the hydrodynamic equations into a closed set is the Chapman-Enskog (ChE) method. In the process it also provides a systematic computation of the transport coefficients from the microscopic theory. Originally it was formulated in the context of classical kinetic theory~\cite{ResiboisBOOK, Dorfman2021}. Here we adapt it to the GHD case. The ChE method assumes the time evolution of $\rho_{\rm p}$ is determined by the hydrodynamic fields, namely $\rho_{\rm p}(\lambda, x | \varrho_{\alpha}(x,t))$, called {\em normal solutions}. With this assumption the GHD equations~\eqref{GHD_full} can be rewritten as
\begin{equation}
	\sum_\alpha\frac{\partial \rho_{\rm p}}{\partial \varrho_{\alpha}} \partial_t \varrho_{\alpha} + \partial_x \left( v_\rho \rho_{\rm p} \right) = \frac{1}{2} \partial_x \left(\mathfrak{D}_{\rho}\partial_x \rho_{\rm p} \right) + \mathcal{I}[\rho_{\rm p}].
\end{equation}
As the external potential does not influence the determination of the transport coefficients we set it here to zero~\cite{longer_paper}. In what follows it will be useful to have a compact notation for the conservation laws. We write $\partial_t \varrho_{\alpha} = D_{\alpha}[\{\varrho_\beta\}]$ where $D_{\alpha}[\{\varrho_\beta\}]$ can be read out from eq.~\eqref{NS_eqns}.
With this we find the ChE equation
\begin{equation} \label{ChE}
	 \sum_{\alpha} \frac{\partial \rho_{\rm p}}{\partial \varrho_{\alpha}} D_{\alpha} + \partial_x \left( v_\rho \rho_{\rm p} \right) = \frac{\delta_\mathfrak{D}}{2} \partial_x \left(\mathfrak{D}_{\rho}\partial_x \rho_{\rm p} \right) + \frac{1}{\delta_\mathcal{I}}\mathcal{I}[\rho_{\rm p}],
\end{equation}
which is an equation for $\rho_{\rm p}$ given the $\{\varrho_\alpha\}$ and contains only spatial derivatives.
The effective velocity and diffusion are now determined fully from the knowledge of $\{\varrho_\alpha\}$. 
In writing eq.~\eqref{ChE} we rescaled $x$ and $t$ by hydrodynamic scales $t_{\rm h}, l_{\rm h}$ introducing two small parameters $\delta_\mathfrak{D}$ and $\delta_\mathcal{I}$ into the equation. They can be understood as generalizations of Knudsen numbers, ie. ratios of characteristic diffusion (collision integral) length-scales to $l_{\rm h}$. 
In ChE method we treat the problem order by order in $\delta$'s writing
\begin{equation}
    \rho_{\rm p} =  \sum_{m,n=0}^\infty \delta_\mathfrak{D}^m \,  \delta_\mathcal{I}^n \, \rho_{\rm p}^{(m,n)},
\end{equation}
from which follow similar expansions for $\mathcal{P}$ and $\mathcal{J}$. Truncating the expansions of $\mathcal{P}, \mathcal{J}$ leads to the closure of \eqref{NS_eqns}. The resulting hydrodynamics depends on the truncation order. At the leading order eq.~\eqref{ChE} gives $\mathcal{I}[\rho_{\rm p}^{(0,0)}]= 0$.
The solution is the locally varying boosted thermal state and this is the {\rm Euler} scale hydrodynamics. Specifically, explicit computations show that the hydrodynamic pressure $\mathcal{P}$ reduces then to the hydrostatic pressure $\mathcal{P}^{(0,0)}=P(\varrho, e)$~\cite{Yang1969,SM, Bouchoule2022}, and the heat current vanish identically $\mathcal{J}^{(0,0)}=0$. The hydrodynamic equations at the Euler scale are
\begin{equation} \label{Euler_eqns}
	\begin{aligned}
	\partial_t \varrho &= - \partial_x (\varrho u), \qquad \partial_t (\varrho u) = - \partial_x (\varrho u^2 + P) , \\
	\partial_t (\varrho e) &= -\partial_x(u \varrho e) - P \partial_x u,
	\end{aligned}
\end{equation}
and are fully determined from the thermodynamics of the system encoded in the equation of state $P = P(\varrho,e)$. We thus partially confirm the phenomenological picture of eq.~\eqref{pheno} in which $\mathcal{P}$ and $\mathcal{J}$, beside the hydrostatic pressure $P$, contain derivative terms which do not enter at the Euler scale. This renders the Euler scale dynamic reversible as can be also witnessed by the conservation of entropy. Indeed, simple computations show that the thermodynamic entropy density $\mathsf{s}(x,t)$ 
obeys the continuity equation $\partial_t \mathsf{s} + \partial_x (v \mathsf{s}) = 0$~\cite{SM}.

At the first orders in both parameters  $\delta_\mathfrak{D},\delta_\mathcal{I}$ we recover the full NS hydrodynamics and find explicit expressions for the transport coefficients.
As announced earlier there are two contributions. The one from the GHD diffusion amounts to evaluating $\mathcal{P}_{\mathfrak{D}}$ and $\mathcal{J}_{\mathfrak{D}}$ on a boosted thermal state. The second contribution is due to the collision integral and is more technically involved. 
The transport coefficients that we find are $\zeta = \zeta_{\mathcal{I}} + \zeta_{\mathfrak{D}}$ and $\kappa = \kappa_{\mathcal{I}} + \kappa_{\mathfrak{D}}$ with
\begin{equation}
\begin{aligned}\label{eq:transportcoeffs}
    \zeta_\mathcal{I} &= \mathscr{h}_1 \mathcal{B} \phi_\zeta, \quad \zeta_{\mathfrak{D}} = \beta \mathscr{h}_1 \mathfrak{D} \mathcal{C} \mathscr{h}_1/2, \\
	\kappa_\mathcal{I} &= \mathscr{h}_2 \mathcal{B} \phi_\kappa,  \quad \kappa_{\mathfrak{D}} = \beta^2 \mathscr{h}_2 \mathfrak{D}\mathcal{C} \mathscr{h}_2/2,
\end{aligned}
\end{equation}
where $\mathcal{B}$ and  $\mathcal{C}$ are kernels of the current and charge susceptibility matrices of the integrable model~\cite{DrudeWeight_LL,DeNardis2022}, explicitly written in the Appendix A.
Additionally, $\beta \equiv 1/T$ and functions $\phi_{\kappa,\zeta}(\lambda)$ are found from the following integral equations
\begin{equation}\label{eq:ChE}
    \Gamma \phi_{\kappa,\zeta}= \chi_{\kappa,\zeta},
\end{equation}
where $\Gamma =- (\delta \mathcal{I}/\delta \rho_{\rm p})\mathcal{C}$ evaluated in a thermal state, and
\begin{equation}
\begin{aligned}
&\chi_\zeta = \sqrt{\beta \varrho} \,  \mathcal{B} h_1-\sqrt{\beta/ \varkappa_T} \, \mathcal{C}(h_0 + \sqrt{c_P/c_V-1} \, h_2), \\
&\chi_\kappa = \beta \sqrt{\varrho c_V}\,  \mathcal{B} ( h_2 - \sqrt{c_P/c_V-1}\,  h_0).
\end{aligned}
\end{equation}
Isothermal compressibility $\varkappa_T$ and specific heats $c_P, c_V$  are those of the unperturbed, integrable model. As we show in~\cite{SM}, thermodynamic quantities are represented as explicit functionals of the underlying thermal state with density $\varrho$ and temperature $T$. 
Functions $h_n(\lambda)$ are charges constructed with  Gram-Schimdt orthonormalization of $ \mathscr{h}_n(\lambda) $ using {\em hydrodynamic scalar product} $(g|h) \equiv g \mathcal{C} h$.
The integral equations \eqref{eq:ChE} are supplemented with additional conditions $(\mathscr{h}_i|\phi_{\kappa,\zeta})=0, \, \, i=0,1,2$, which render their solutions unique. Equations \eqref{eq:ChE} naturally generalize ChE integral equations~\cite{ResiboisBOOK} to interacting, nearly integrable models.
In the process we confirm that there are no other gradient terms in $\mathcal{P}$ and $\mathcal{J}$ beside the two given in the phenomenological expression~\eqref{pheno}. 

Both contributions to the transport coefficients have a feature known from the standard kinetic theory~\cite{ResiboisBOOK, Resibois1970}. Namely, they are a simple generalization of coefficients found in a linearized hydrodynamics. Therefore, in what follows we focus on the linearized theory, which also has the advantage of revealing the physical ingredients and defer the further computations within the ChE method to a longer publication~\cite{longer_paper}.

%==============================================================================
{\bf Linearized hydrodynamics:} We linearize the GHD equations~\eqref{GHD_full} around the homogeneous thermal state such that $\rho_{\rm p}(x,t, \lambda) = \rho_{\rm th}(\lambda) + \delta \rho_{\rm p}(x, t, \lambda)$,
\begin{equation} \label{GHD_linear}
	\partial_t \delta\rho_{\rm p} + \mathcal{A}_{\rm th} \partial_x  \delta\rho_{\rm p} = \frac{1}{2} \mathfrak{D}_{\rm th} \partial_x^2 \delta\rho_{\rm p} +  \left(\frac{\delta \mathcal{I}}{\delta \rho_{\rm p}}\right)_{\rm th} \delta\rho_{\rm p},
\end{equation}
where $\mathcal{A}$ is the kernel of flux Jacobian matrix~\cite{DrudeWeight_LL,lecture_notes_GHD}. The linearized GHD equation~\eqref{GHD_linear} leads to an eigenequation for perturbations of conserved charges $\delta q_n(x,t)  = \int {\rm d}\lambda\, h_n(\lambda) \delta \rho_{\rm p}(\lambda; x,t)$, for more details see the Appendix C. We then consider a time evolution of a single mode $\delta q_n(x,t) \sim \exp(i k x - \Lambda(k) t) \delta q_n(k)$.
For small $k$ a standard perturbation theory methods apply and we find two sound modes and a heat mode (thermal mode) with dispersions (we neglect $\mathcal{O}(k^3)$ terms)
\begin{equation}\label{eq:modes}
\begin{aligned}
	\Lambda_{\pm}(k) &= \pm i v_s k +\frac{1}{2\varrho} \bigg[ \left( \frac{1}{c_V}-\frac{1}{c_P} \right) \kappa + \zeta \bigg] k^2 , \\
	\Lambda_{\rm th}(k) &= \frac{ \kappa}{\varrho c_P}k^2,
\end{aligned}
\end{equation}
which agree with dispersion relations of linearized NS equations. The sound velocity 
is equal to thermodynamic formula $v_s = 1/\sqrt{\varrho \varkappa_S}$, where $\varkappa_S$ is the adiabatic compressibility of the integrable model calculated for the thermal background state $\rho_{\rm th}$. 
The transport coefficients $\zeta$ and $\kappa$ are given by~\eqref{eq:transportcoeffs} computed with the same state $\rho_{\rm th}$. The details are referred to~\cite{SM}.

The corresponding eigenmodes are related to the hydrodynamic fields $\varrho$, $u$ and $T$
\begin{equation}
\begin{aligned}
	\delta \varrho = \varrho \sqrt{T \varkappa_T}\, \delta q_0, \, \, \, \, \delta u = \sqrt{T/\varrho}\, \delta q_1, \, \, \, \,
	\delta T = T/\sqrt{\varrho c_V}\, \delta q_2, 
\end{aligned}
\end{equation}
where we took $\varrho(x,t) = \varrho + \delta \varrho(x,t)$, $u(x,t) = 0 + \delta u(x,t)$ and $T(x,t) = T + \delta T(x,t)$. Therefore the first $3$ orthonormalized charges are equivalent to the hydrodynamic degrees of freedom. Expressing heat and sound modes in terms of $\delta \varrho, \delta u, \delta T$ we recover precisely the  modes of NS equations~\cite{SM}.

%==============================================================================
{\bf Application to the Lieb-Liniger model:}
We now apply the introduced formalism to a situation relevant to experiments with cold atomic gases. We consider a setup of two 1d Bose gases, described by the integrable Lieb-Liniger (LL) model~\cite{Lieb1963a}, coupled by a long-range interaction which breaks the integrability. The integrable part of the Hamiltonian is $H_{\rm int} = H_1 + H_2$ with
\begin{equation}
	H_i = \int {\rm d}x\, \psi_i^{\dagger}(x) \left(-\frac{1}{2}\partial_x^2 +  \frac{c}{2}\, \psi_i^{\dagger}(x) \psi_i(x)\right) \psi_i(x),
\end{equation}
where we work in the units $\hbar = m = 1$ and $\psi_i(x), \psi_i^{\dagger}(x)$ are canonical bosonic fields. The integrability breaking coupling between the tubes is given by
$V_2 = \int{\rm d}x {\rm d}y V(x-y) \hat{\rho}_1(x) \hat{\rho}_2(y)$, 
with $\hat{\rho}_i(x) = \psi_i^{\dagger}(x) \psi_i(x)$ and $V(x-y)$ interaction potential. In an experimental realizations with cold-atomic gases it originates from dipole-dipole interactions and its strength is tunable~\cite{Tang2018,Li2023}.

Of special interest is the case when the two tubes are initially in the same state. Then they stay the same throughout the dynamics and the system can be described by a single equation~\eqref{GHD_full}. The collision integral based on form-factor approach~\cite{DeN2015,DeN2018,Panfil2021} and the resulting relaxation for the homogeneous initial state was studied in~\cite{PGK}. For the computation of the transport coefficients we need $\mathcal{I}$ linearized around a thermal state, c.f.\eqref{eq:ChE}. The derivation is presented in~\cite{SM}. The result for the matrix elements of $\Gamma$ in the basis $\{h_j\}$~is
\begin{align}
	\Gamma_{nm}\! = \frac{2 \pi}{\tau} \!\int {\rm d}\boldsymbol{\lambda}_1 {\rm d}\boldsymbol{\lambda}_2\, g_{nm}(\lambda_1, \lambda_2) \left(\mathcal{T}^{\rm dr}(\lambda_1, \lambda_2)\right)^2 |v_{12}|,
\end{align}
with the integration measure ${\rm d}\boldsymbol{\lambda} = \rho_{\rm p}(\lambda) (1-n(\lambda)) {\rm d}\lambda$, the filling function $n(\lambda)$, $v_i = v(\lambda_i)$, $v_{12} = v_1 - v_2$, and
\begin{equation}
	g_{nm}(\lambda_1, \lambda_2) =  \int {\rm d}\boldsymbol{\lambda}\, (k'(\lambda))^2 \frac{ \mathcal{S}[h_n^{\rm Dr}]  \mathcal{S}[h_m^{\rm Dr}]}{(v - v_{1})^2(v - v_{2})^2}.
\end{equation}
Here $\mathcal{T}$ is the differential scattering kernel, for the LL model $\pi \mathcal{T}(\lambda-\lambda') = c/(c^2 + (\lambda-\lambda')^2)$ and ${\rm dr}$ and ${\rm Dr}$ are dressing operations~\cite{PhysRevLett.124.140603}, $k(\lambda)$ is the Dressed momentum while $S[\varphi] \equiv S[\varphi](\lambda, \lambda_1, \lambda_2)$ is 
\begin{equation}
    S[\varphi] = \frac{\varphi'}{k'(v_1 - v_2)} + \frac{\varphi_1'}{k_1'(v_2 - v)} + \frac{\varphi_2'}{k_2'(v - v_1)}.
\end{equation}
The timescale $\tau$ is related to the strength of inter-tube coupling $\tau^{-1}=\tilde{V}^2m/(4 \hbar^3)$, where $\tilde{V}^2= \int dk \tilde{V}^2(k) k^2$ and $\tilde{V}(k)$ is Fourier transform of $V(x)$.

\begin{figure}
    \centering
    \includegraphics[scale=0.55]{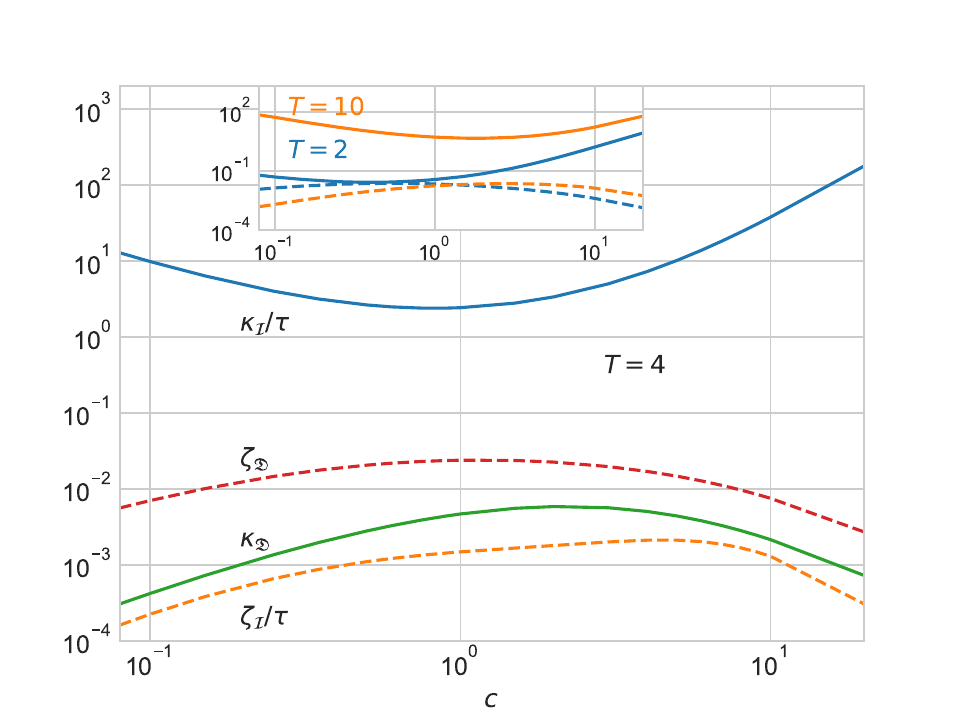}
    \caption{Transport coefficients in the log-log scale for the coupled LL models as a function of intratube interaction coupling $c$. In the main plot we show separate contributions to the thermal conductivity and viscosity from integrable and non-integrable collisions. In the inset we show the total thermal conductivity and viscosity for $\tau = 10^{-2}$. The viscosity is non-zero across different values of temperature and couplings.}
    \label{fig:transport}
\end{figure}

The transport coefficients are then determined from~\eqref{eq:transportcoeffs} and the numerical solution is shown in Fig.~\ref{fig:transport}. We focus on the strong coupling regime of the LL model in the nomenclature of~\cite{PhysRevLett.91.040403} where both interaction and thermal fluctuations are important. We find that there are two competing behaviors. When intratube interactions dominate the resulting fluid has relatively small viscosity and large thermal conductivity. Instead, when intertube interactions dominate then both viscosity and thermal conductivity are of the same order. In the limit of noninteracting integrable models, $c \to 0$ or $c^{-1} \to 0$, the bulk viscosity vanishes and thermal conductivity diverges~\cite{SM} as expected from the classical kinetic theory of weakly interacting models~\cite{Balescu1975}.

%==============================================================================
{\bf Outlook:} In this work we have shown that the NS equations universally arise for $1d$ quantum systems with weakly broken integrability and with Galilean invariance, assuming the GHD equation supplemented with the collision integral. We have provided an explicit equations~\eqref{eq:ChE} for the transport coefficients linking the effective hydrodynamic description with the exact microscopic thermodynamics of the integrable model. Unlike the standard approach in the kinetic theory, we treat an important part of the interactions non-perturbatively. One consequence of this is a non-zero viscosity, which in the derivation of the NS equations based on the free theory, is identically zero~\cite{ResiboisBOOK,spohn2012large}. 
The resulting fluid has two regimes. One dominated by the integrability breaking collisions with relatively small viscosity, and the second, dominated by the integrable collisions with thermal conductivity and viscosity of the same order.

It is well-known that transport in one dimension with conserved momentum becomes anomalous upon introducing noise to the system~\cite{Beijeren2012,Narayan2002,Forster1977,Spohn2014}. The linearized framework with $U(x)=0$ presented here together with initial results for GHD with noise~\cite{Bastianello2020} open possibility for further explorations in this direction.

Recently, the NS equations were also found for the low temperature dynamics of the GHD without an explicit integrability breaking terms~\cite{PhysRevLett.132.243402}. The effective projection to the three lowest conserved densities is attributed there to the low temperatures and holds at intermediate timescales. The transport coefficients are then of the form $\zeta_{\mathfrak{D}}$ and $\kappa_{\mathfrak{D}}$ taken in the $T \to 0$ limit, namely they origin from the integrable collisions.

Our results apply also to classical integrable Galilean-invariant models, whose dynamics is described by~\eqref{GHD_full}. In such cases collision integrals may follow from BBGKY equations~\cite{biagetti2024BBGKY} or be constructed phenomenologically, as in the relaxation time approximation~\cite{Lopez2021}. The methods introduced here can be also directly generalized to models without the Galilean invariance, e.g. spin chains.

%==============================================================================
The authors thank Romain Vasseur for his inspiration and encouragement to tackle this problem. We also thank Alvise Bastianello and Sarang Gopalakrishnan for discussions.
The authors acknowledge support by the National Science Centre (NCN), Poland via projects 2018/31/D/ST3/03588 and 2022/47/B/ST2/03334.
\newline
\textit{Appendix A: Integral kernels--} For a linear operator $\mathcal{O}$ we introduce the following notation for its action on arbitrary functions $\psi,\varphi$
\begin{align}
    (\mathcal{O} \psi)(\lambda) &= \int {\rm d} \mu \mathcal{O}(\lambda,\mu) \psi(\mu), \\
\psi\mathcal{O}\varphi &= \int {\rm d} \lambda {\rm d} \mu \mathcal{O}(\lambda,\mu) \psi(\lambda) \varphi(\mu),
\end{align}
and use the standard notation for kernel multiplication
\begin{equation}
    (\mathcal{O}_1 \mathcal{O}_2)(\lambda,\mu) = \int {\rm d} \nu \mathcal{O}_1(\lambda,\nu) \mathcal{O}_2(\nu,\mu).
\end{equation}
The kernels for operators $\mathcal{A}, \mathcal{B}, \mathcal{C}$ appearing in the main text are as follows
\begin{align}
    \mathcal{A} &= (1-n \mathcal{T})^{-1} v (1-n \mathcal{T}), \\
    \mathcal{B} &= (1-n \mathcal{T})^{-1} \rho_{\rm p} f v (1-\mathcal{T}n)^{-1}, \\
    \mathcal{C} &= (1-n \mathcal{T})^{-1} \rho_{\rm p} f (1-\mathcal{T}n)^{-1}.
\end{align}
where $f$ is the statistical factor~\cite{lecture_notes_GHD} (for fermionic statistics  $f=1-n$). In the definitions above $\rho_{\rm p},n, f,v$ should be viewed as diagonal kernels and we have $(\mathcal{T}\psi)(\lambda)=\int {\rm d} \lambda'\mathcal{T}(\lambda-\lambda') \psi(\lambda')$. Lastly, the diffusion kernel is 
\begin{equation}
	\mathfrak{D} = (1 - n \mathcal{T})^{-1} \rho_{\rm tot}^{-1} \tilde{\mathfrak{D}} \rho_{\rm tot}^{-1}  (1 - n\mathcal{T}), 
\end{equation}
with $\tilde{\mathfrak{D}}(\theta, \alpha) = \delta(\theta - \alpha) w(\theta) - W(\theta, \alpha)$ and
\begin{align}
	W(\theta, \alpha) =  \rho_{\rm p}(\theta) f(\theta) \left(\mathcal{T}^{\rm dr}(\theta, \alpha)\right)^2 | v(\theta) - v(\alpha)|, 
\end{align}
where $w(\theta) =  \int {\rm d}\alpha W(\alpha, \theta)$ and $\mathcal{T}^{\rm dr}(\lambda,\lambda')$ fulfills
\begin{equation}
\label{eqSM:Tdressed}
    \mathcal{T}^{\rm dr}(\lambda,\lambda')= \mathcal{T}(\lambda-\lambda') + \int {\rm d} \lambda'' \mathcal{T}(\lambda-\lambda'')n(\lambda'') \mathcal{T}^{\rm dr}(\lambda'',\lambda').
\end{equation}
\newline
\newline
\textit{Appendix B: Boosted thermal states in the Thermodynamic Bethe Ansatz--} We discuss here the basics of Thermodynamic Bethe Ansatz (TBA)\cite{Mossel2012,2012PhRvL.109q5301C} for a model with fermionic statistics (for generic case see~\cite{SM}) and give a precise meaning of boosted thermal states within this framework. The result of the TBA is that averages (of local operators or products thereof) in the grand canonical ensemble or generalized Gibbs ensemble are thermodynamically equal to expectation values computed in an eigenstate with rapidity distribution $\rho_{\rm p}$. This rapidity distribution follows from the solving the TBA equations that we now summarize.  We start with bare pseudo-energy $\epsilon_0$ which takes the following form
\begin{equation} \label{pseudo_TBA}
	\epsilon_0(\lambda, x, t) = \sum_n \mathscr{h}_n(\lambda) \beta_n.
\end{equation}
The pseudo-energy $\epsilon$ and total density of states $\rho_{\rm tot}$ is then determined from the generalized TBA equations
\begin{align}
	\epsilon(\lambda) &= \epsilon_0(\lambda) - \int {\rm d}\mu\, \mathcal{T}(\lambda - \mu) \log\left(1 + e^{-\epsilon(\mu)}\right), \\
	\rho_{\rm tot}(\lambda) &=\frac{1}{2 \pi} + \int {\rm d}\mu\, \mathcal{T}(\lambda - \mu) n(\mu) \rho_{\rm tot}(\mu).
\end{align}
Solving the first equation yields $\epsilon(\lambda)$ which enters the filling function 
\begin{equation}
	n(\lambda) = \frac{1}{1 + e^{\epsilon(\lambda)}},
\end{equation}
and in turn allows to solve the second equation. The total density $\rho_{\rm tot}$ determines then the particles' density according to $\rho_{\rm p}(\lambda) = n(\lambda) \rho_{\rm tot}(\lambda)$. 

For thermal state $\bar{\rho}_{\rm p}(\lambda)$ with chemical potential $\mu$ and temperature $T$ one has $\bar{\beta}_0=-\mu/T, \, \bar{\beta}_1=0, \, \bar{\beta}_2 =1/T, \, \bar{\beta}_{n>2}=0$. The Galilean invariance implies that after boosting by velocity $u$ the state is a TBA state with $\rho_{\rm p}(\lambda) = \bar{\rho}_{\rm p}(\lambda - u)$ and with Lagrange multipliers $\beta_0 = \bar{\beta}_0 + \frac{1}{2}u^2 \bar{\beta}_2, \,  \beta_1 =-u\bar{\beta}_2, \,  \beta_2=\bar{\beta}_2, \, \beta_{n>2}=0$.
\newline
\newline
\textit{Appendix C: Further informations on the linearized equations--}We elaborate here on the computations involving the linearized Boltzmann equation. As indicated in the main text the structure of the resulting equations is simpler if we work in the orthonormal basis of the conserved charges with respect to the hydrodynamic scalar product $(g|h)$. Standard computations using the methods of Thermodynamic Bethe Ansatz yield, see~\cite{SM},
\begin{equation} \label{linear_GHD}
	\partial_t \delta q_n + A_{nm} \partial_x \delta q_m -D_{nm} \partial_x^2 \delta q_m = - \Gamma_{nm}  \delta q_m,
\end{equation}
where we assume a summation over the repeated indices. Moreover, $A_{nm}$, $D_{nm}$ and $\Gamma_{nm}$ are now interpreted as matrix elements of operators associated with effective velocity, diffusion and collision integral evaluated in the orthonormal basis. We observe that $D_{0n} = D_{n0}=0$ which is a consequence of Markovianity of the GHD diffusion operator~\cite{hydro_diff_prl,SM}, while $\Gamma_{jn} = 0$ for $j,n=0,1,2$ which reflects the collision invariants~\eqref{coll_inv}.
The linearized GHD equations~\eqref{linear_GHD} can be now solved perturbatively in gradients of $\delta q_n$. This is the easiest to formulate by going to the Fourier space and considering a time evolution of a single mode $\delta q_n(x,t) \sim \exp(i k x - \Lambda(k) t) \delta q_n(k)$. This leads to the eigenvalue problem
\begin{equation}\label{eq:eigenproblem}
	\left(\Gamma_{nm} + i k A_{nm} +  k^2 D_{nm}\right) \delta q_{m}(k) = \Lambda(k) \delta q_{n}(k),
\end{equation}
with $\Lambda(k)$ the dispersion relations for different modes. Among them there are $3$ gapless modes related to the $3$ conservation laws respected by the collision integral. The problem can be treated analytically in small $k$ limit using standard perturbation theory around the spectrum of $\Gamma$ operator. The three zero modes of $\Gamma$ yield gapless dispersion relations \eqref{eq:modes} of heat and sound modes. At the first order in $k$ the only non-trivial coefficient is sound velocity $v_s = \left(A_{1,0}^2 + A_{2,1}^2\right)^{1/2}$, which agrees with the standard thermodynamic formula $v_s = 1/\sqrt{\varrho \varkappa_S}$, see~\cite{SM}. At the second order in $k$ after lengthy calculation we recover subleading corrections as in~\eqref{eq:modes} with transport coefficients which are determined from \eqref{eq:transportcoeffs}. These derivations are presented in~\cite{SM}. 

\let\oldaddcontentsline\addcontentsline% Store \addcontentsline
\renewcommand{\addcontentsline}[3]{}% Make \addcontentsline a no-op
\bibliography{biblio}
\let\addcontentsline\oldaddcontentsline% Restore \addcontentsline

\clearpage 
\pagebreak

\onecolumngrid
\begin{center}
\textbf{\large Supplemental Material:\\
Navier-Stokes equations for nearly integrable quantum gases }\\[5pt]
Maciej Łebek and Miłosz Panfil \\

{\small \sl Faculty of Physics, University of Warsaw, Pasteura 5, 02-093 Warsaw, Poland}\\
\end{center}
\setcounter{equation}{0}
\setcounter{figure}{0}
\setcounter{table}{0}
\setcounter{page}{1}
\setcounter{section}{0}
\numberwithin{equation}{section} 
\makeatletter

\vspace{0cm}

\onecolumngrid
In the Supplemental Material we provide computations supporting the results presented in the main text. 

In Sec.~\ref{secSM:sec1} we introduce the basic ingredients of the GHD equation: effective velocity, diffusion kernel and susceptibility matrices along with properties of these objects relevant in the present context. In Sec.~\ref{secSM:sec2} we derive hydrodynamic conservation laws for the three conserved charges and introduce dynamic pressure and heat current. Next, in Sec.~\ref{secSM:sec3} we analyze the basic properties of thermal boosted state of integrable models. We show that for such states the heat current vanishes and dynamic pressure becomes the hydrostatic pressure. In Sec.~\ref{secSM:sec4} we derive expressions for thermodynamic quantities such as sound velocity and specific heats in terms of hydrodynamic matrices. 

The following sections concern the linearized hydrodynamics and computation of transport coefficients. First, in Sec.~\ref{secSM:sec5} we present a derivation of a linearized transport equation for local charge densities. In Sec.~\ref{secSM:sec6} we linearize Navier-Stokes equations and find the dispersion relations of heat and sound modes. We use these results in Sec.~\ref{secSM:sec7} where we treat the eigenproblem perturbatively and identify expressions for bulk viscosity and thermal conductivity. Additionally, we give  numerical results for the transport coefficients. Later in Sec.~\ref{secSM:sec8} we show that the results found in the perturbation theory can be formulated as generalized Chapman-Enskog integral equations, which are stated in the main text. 

Finally, in Sec.~\ref{secSM:sec9} we analyze the collision integral for two Lieb-Liniger gases coupled by the density-density interactions.

\setcounter{secnumdepth}{2}
\tableofcontents

\section{Effective velocity, diffusion operator and susceptibility matrices}\label{secSM:sec1}

We consider a Galilean-invariant model with energy $E(\lambda)=\frac{1}{2} \lambda^2$ and momentum $p(\lambda) = \lambda$. We set the mass to unity. As in the main text we introduce functions $\mathscr{h}_j(\lambda)=\lambda^j/j!$ related to ultra-local charges.
The effective velocity \cite{lecture_notes_GHD} is given by $v(\lambda)=(E')^{\rm dr}(\lambda)/ (p')^{\rm dr}(\lambda)$ where the dressing operation of an arbitrary function $g(\lambda)$ is implemented by the following integral equation
\begin{equation}
\label{eqSM:dressing}
	g^{\rm dr}(\lambda) = g(\lambda) + \int {\rm d}\lambda' \mathcal{T}(\lambda- \lambda')n(\lambda') g^{\rm dr}(\lambda'),
\end{equation}
and $\mathcal{T}(\lambda-\lambda')$ is a symmetric scattering kernel, which depends on the model. For instance, in the case of Lieb-Liniger model~\cite{Lieb1963a} it reads $\mathcal{T}(\lambda)=\frac{c}{\pi} \frac{1}{c^2+\lambda^2}$, where $c$ is the coupling constant. We also defined the occupation function $n(\lambda)=\rho_{\rm p}(\lambda)/ \rho_{\rm tot}(\lambda)$, where total density of states $\rho_{\rm tot}(\lambda)$ is 
\begin{equation}
\label{eqSM:rhotot}
    \rho_{\rm tot}(\lambda) = \frac{1}{2 \pi} + \int \dd \lambda' \mathcal{T}(\lambda-\lambda') \rho_{\rm p}(\lambda).
\end{equation}
It can be shown that the effective velocity fulfils the following integral equation~\cite{lecture_notes_GHD}
\begin{equation}
\label{eqSM:vinteq}
    v(\lambda) = \mathscr{h}_1(\lambda) + 2 \pi \int \dd \lambda' \mathcal{T}(\lambda-\lambda') \rho_{\rm p}(\lambda')\left(v(\lambda')-v(\lambda)\right).
\end{equation}
With this, it is straightforward to check the identity used in the main text
\begin{equation}
\label{eqSM:veffidentity}
    \int \dd \lambda v(\lambda) \rho_{\rm p}(\lambda) = \int \dd \lambda \mathscr{h}_1(\lambda) \rho_{\rm p}(\lambda).
\end{equation}
Indeed, we have
\begin{equation}
    \int \dd \lambda v(\lambda) \rho_{\rm p}(\lambda) = \int d \lambda \mathscr{h}_1(\lambda) \rho_{\rm p}(\lambda) +2\pi \int \dd \lambda \dd \lambda' \mathcal{T}(\lambda-\lambda') \rho_{\rm p}(\lambda) \rho_{\rm p}(\lambda') \left(v(\lambda')-v(\lambda)\right)= \int \dd \lambda \mathscr{h}_1(\lambda) \rho_{\rm p}(\lambda),
\end{equation}
where the integral $\int \dd \lambda \dd \lambda' (\ldots)$ vanished due to the symmetry of the kernel $\mathcal{T}(-\lambda)=\mathcal{T}(\lambda)$.

In addition to the Euler-scale term involving the effective velocity, there is also the diffusion term~\cite{hydro_diff_prl} $\frac{1}{2}\partial_x \left (\mathfrak{D}\partial_x \rho_{\rm p} \right)$ with $(\mathfrak{D}\partial_x \rho_{\rm p})(\theta)= \int \dd \alpha \mathfrak{D}(\theta,\alpha)\partial_x \rho_{\rm p}(\alpha)$. The diffusion kernel is 
\begin{equation}
	\mathfrak{D} = (1 - n \mathcal{T})^{-1} \rho_{\rm tot}^{-1} \tilde{\mathfrak{D}} \rho_{\rm tot}^{-1}  (1 - n\mathcal{T}), \qquad \tilde{\mathfrak{D}}(\theta, \alpha) = \delta(\theta - \alpha) w(\theta) - W(\theta, \alpha),
\end{equation}
and
\begin{align}
	W(\theta, \alpha) =  \rho_{\rm p}(\theta) f(\theta) \left(\mathcal{T}^{\rm dr}(\theta, \alpha)\right)^2 | v(\theta) - v(\alpha)|, \qquad w(\theta) =  \int {\rm d}\alpha W(\alpha, \theta),
\end{align}
where $f(\theta)$ denotes statistical factor~\cite{lecture_notes_GHD} defined in  \eqref{eqSM:f_factor}, which depends on the particles' statistics. The dressed scattering kernel fulfills
\begin{equation}
\label{eqSM:Tdressed}
    \mathcal{T}^{\rm dr}(\lambda,\lambda')= \mathcal{T}(\lambda-\lambda') + \int \dd \lambda'' \mathcal{T}(\lambda-\lambda'')n(\lambda'') \mathcal{T}^{\rm dr}(\lambda'',\lambda').
\end{equation}
Note also that our definition of $\tilde{\mathfrak{D}}$ differs from the standard one~\cite{hydro_diff_prl} by a factor of $\rho_{\rm tot}^2(\theta)$.

From the Galilean invariance follows that diffusion kernel has left eigenvector with zero eigenvalue~\cite{hydro_diff_prl},
\begin{equation}
\int  {\rm d} \theta \mathfrak{D}(\theta,\alpha)=0.
\end{equation} 
This implies that the local density of particles is conserved by the diffusion term. Moreover from Markovianity of the diffusion kernel, there is a corresponding right eigenvector with zero eigenvalue, namely
\begin{equation}
    \int {\rm d} \alpha \mathfrak{D}(\theta,\alpha) (\mathcal{C}\mathscr{h}_0)(\alpha)=0,
\end{equation}
where $\mathcal{C}$ is the charge susceptibility matrix defined as~\cite{lecture_notes_GHD, DrudeWeight_LL}
\begin{equation}
\label{eqSM:Ckernel}
    \mathcal{C} = (1-n \mathcal{T})^{-1} \rho_{\rm p} f (1-\mathcal{T}n)^{-1}.
\end{equation} 
We have used here the fact~\cite{lecture_notes_GHD} that $\rho_{\rm tot}(\lambda)=\mathscr{h}_0^{\rm dr}(\lambda)/(2 \pi)$. For later use, we also introduce here the current susceptibility matrix $\mathcal{B}$~\cite{lecture_notes_GHD, DrudeWeight_LL}
\begin{equation}
\label{eqSM:Bkernel}
    \mathcal{B} = (1-n \mathcal{T})^{-1} \rho_{\rm p} f v (1-\mathcal{T}n)^{-1},
\end{equation} 
we will discuss \textit{matrix elements} of the operators $\mathcal{B}$ and $\mathcal{C}$ in more detail in Sec. \ref{secSM:sec4}.

\section{Hydrodynamic conservation laws}\label{secSM:sec2}

We derive here the conservation laws for the 3 hydrodynamic modes. We adopt notation from the main text for the expectation values of charges
\begin{equation}
    \mathscr{q}_n= \int \dd \lambda \mathscr{h}_n(\lambda) \rho_{\rm p}(\lambda).
\end{equation}
We start with the GHD equation including the diffusion, force term $\mathfrak{f}= -\partial_xU(x)$ and the Boltzmann scattering integral
\begin{equation} \label{eqSM:GHD_full}
	\partial_t \rho_{\rm p}  + \partial_x \left(v_{\rho} \rho_{\rm p}  \right) + \mathfrak{f} \partial_\lambda \rho_{\rm p} = \frac{1}{2} \partial_x \left(\mathfrak{D}_{\rho}\partial_x \rho_{\rm p}  \right) + \mathcal{I}[\rho_{\rm p} ].
\end{equation}
To obtain the conservation laws we multiply the GHD equation by $\mathscr{h}_j(\lambda)$ and integrate over $\lambda$. 
\begin{itemize}
\item Mass conservation: for $j=0$ we obtain
\begin{equation}
	\partial_t \langle \mathscr{q}_0\rangle  + \partial_x \langle \mathscr{q}_1\rangle = 0,
\end{equation}
where we used that both the diffusion and the collision term conserve the particles' density. This equation reflects the conservation of particles' density. We have also used that
\begin{equation}
	\int {\rm d}\lambda\, v(\lambda) \rho_{\rm p} (\lambda) = \int {\rm d}\lambda\, \mathscr{h}_1(\lambda) \rho_{\rm p} (\lambda),
\end{equation}
\item Momentum conservation: for $j=1$ we obtain
\begin{equation}
	\partial_t \langle \mathscr{q}_1 \rangle + \partial_x \langle \mathscr{q}_1 v \rangle =- \partial_x \mathcal{P}_\mathfrak{D} +\mathfrak{f} \varrho,
\end{equation}
where we defined
\begin{equation}
	\mathcal{P}_\mathfrak{D} =- \frac{1}{2} \int {\rm d}\lambda {\rm d}\mu \, \mathscr{h}_1(\lambda) \mathfrak{D}_{\rho} (\lambda, \mu)\partial_x \rho_{\rm p}(\mu).
\end{equation}
The average in the second term can be decomposed into two contributions: the convective flow and the pressure term. To this end we define the velocity field
\begin{equation}
\label{eqSM:udef}
	u = \frac{\langle \mathscr{q}_1 \rangle}{\langle \mathscr{q}_0 \rangle}, 
\end{equation}
and introduce rapidity variable $\xi$ defined with respect to $u$, $\lambda = u + \xi$. From the definition of $u$ it then follows that
\begin{equation}
	\int {\rm d}\xi\, \xi \rho_{\rm p}(u + \xi) = 0,
\end{equation}
which is easily shown by transforming back to rapidity $\lambda$. We then have
\begin{align}
	\langle \mathscr{q}_1 v \rangle = \int {\rm d}\lambda v(\lambda) \lambda \rho_{\rm p}(\lambda) = \int {\rm d}\xi (u+\xi) v(u+\xi)\rho_{\rm p}(u + \xi) = \frac{\langle \mathscr{q}_1 \rangle ^2}{\langle \mathscr{q}_0 \rangle} + \mathcal{P}_v,
\end{align}
where we defined the pressure by
\begin{equation}
	\mathcal{P}_v = \int {\rm d}\xi\, \xi v(u+\xi)  \rho_{\rm p}(u + \xi) = \int {\rm d}\lambda\, (\lambda - u) v(\lambda) \rho_{\rm p}(\lambda).
\end{equation}
We will later see that for a system locally in the thermal equilibrium state this reduces to the hydrostatic pressure $P$. 

Summing up, the continuity equation reads
\begin{equation}
	\partial_t \langle \mathscr{q}_1 \rangle + \partial_x \left(\frac{\langle \mathscr{q}_1 \rangle^2}{\langle \mathscr{q}_0 \rangle} + \mathcal{P} \right) = \mathfrak{f} \varrho,
\end{equation}
with $\mathcal{P} = \mathcal{P}_v + \mathcal{P}_\mathfrak{D}$.

\item Energy conservation: Finally, for $j=2$ we find
\begin{equation}
\label{eqSM:balance_ene}
	\partial_t \langle \mathscr{q}_2 \rangle + \partial_x \langle \mathscr{q}_2 v \rangle = -\partial_x \left( u\mathcal{P}_\mathfrak{D} +  \mathcal{J}_\mathfrak{D} \right)+\mathfrak{f}\varrho u,	
\end{equation}
where we have defined
\begin{equation}
	u \mathcal{P}_\mathfrak{D}+\mathcal{J}_\mathfrak{D} = -\frac{1}{2} \int {\rm d}\lambda {\rm d}\mu\, \mathscr{h}_2(\lambda) \mathfrak{D}_{\rho}(\lambda, \mu) \partial_x \rho_{\rm p}(\mu).
\end{equation}
The average on the left hand side of \eqref{eqSM:balance_ene} can be again divided into various contributions. Transforming to $\xi$ variable we have
\begin{equation}
	\langle \mathscr{q}_2 v \rangle = \frac{1}{2}\int {\rm d}\xi (u + \xi)^2 v(u + \xi) \rho_{\rm p}(u + \xi) = \frac{1}{2} u^2 \langle \mathscr{q}_1 \rangle + u \mathcal{P}_v + \frac{1}{2}\int {\rm d}\xi \xi^2 v(u + \xi) \rho_{\rm p}(u + \xi).
\end{equation}
The remaining integral can be still divided into two contributions according to
\begin{equation}
	\frac{1}{2}\int {\rm d} \xi \xi^2 v(u + \xi) \rho_{\rm p}(u + \xi) = \frac{1}{2}u \int {\rm d}\xi\, \xi^2 \rho_{\rm p}(u + \xi) + \frac{1}{2}\int {\rm d}\xi\, \xi^2 (v(u + \xi) - u)\rho_{\rm p}(u + \xi).
\end{equation}
The first integral, by going back to $\lambda$ variable, becomes
\begin{equation}
	 \frac{1}{2}\int {\rm d}\xi\, \xi^2 \rho_{\rm p}(u + \xi) = \langle \mathscr{q}_2\rangle - \frac{1}{2}\frac{\langle \mathscr{q}_1 \rangle^2}{\langle \mathscr{q}_0 \rangle},
\end{equation}
whereas the second defines the $\mathcal{J}_v$ contribution to the heat current
\begin{equation}
	\mathcal{J}_v = \frac{1}{2} \int {\rm d}\xi \xi^2 (v(u+\xi)-u) \rho_{\rm p}(u+\xi) = \frac{1}{2} \int {\rm d}\lambda (\lambda - u)^2 (v(\lambda)-u) \rho_{\rm p}(\lambda).
\end{equation}
The total heat current is $\mathcal{J} = \mathcal{J}_v + \mathcal{J}_\mathfrak{D}$. 
Finally we obtain
\begin{equation}
	\partial_t \langle \mathscr{q}_2 \rangle + \partial_x \left(\frac{\langle \mathscr{q}_1 \rangle \langle \mathscr{q}_2 \rangle }{\langle \mathscr{q}_0 \rangle} + \frac{\langle \mathscr{q}_1 \rangle }{\langle \mathscr{q}_0 \rangle} \mathcal{P} + \mathcal{J}\right) = \mathfrak{f} \varrho u.
\end{equation}
\end{itemize}
Summarizing we obtain the following set of equations
\begin{align}
	&\partial_t \langle \mathscr{q}_0\rangle  + \partial_x \langle \mathscr{q}_1\rangle = 0, \qquad \partial_t \langle \mathscr{q}_1 \rangle + \partial_x \left(\frac{\langle \mathscr{q}_1 \rangle^2}{\langle \mathscr{q}_0 \rangle} + \mathcal{P} \right) = \mathfrak{f} \varrho, \\
	&\partial_t \langle \mathscr{q}_2 \rangle + \partial_x \left(\frac{\langle \mathscr{q}_1 \rangle \langle \mathscr{q}_2 \rangle }{\langle \mathscr{q}_0 \rangle} + \frac{\langle \mathscr{q}_1 \rangle }{\langle \mathscr{q}_0 \rangle} \mathcal{P} + \mathcal{J}\right) = \mathfrak{f} \varrho u.
\end{align}

The continuity equations written above are for moments of the particles distribution. More standard form is found if we express them in terms of the velocity field $u$ and internal energy field $e$ defined by
\begin{equation}
\label{eqSM:edef}
	e = \frac{\langle \mathscr{q}_2 \rangle}{\langle \mathscr{q}_0 \rangle} - \frac{1}{2} u^2.
\end{equation}
Introducing also $\varrho(x,t) = \langle \mathscr{q}_0 \rangle$ we find
\begin{align}
	&\partial_t \varrho  + \partial_x (\varrho u ) = 0, \qquad \partial_t (\varrho u)  + \partial_x \left( \varrho u^2 + \mathcal{P} \right) = \mathfrak{f} \varrho, \\
	&\partial_t \left( \varrho e + \frac{1}{2} u^2 \varrho \right) + \partial_x \left(u \varrho e + \frac{1}{2}u^3 \varrho +  u \mathcal{P} + \mathcal{J}\right) = \mathfrak{f} \varrho u.
\end{align}
This can be still simplified using that
\begin{equation}
	\frac{1}{2}\partial_t \left(u^2 \varrho \right) = \frac{1}{2} u^2 \partial_x \left(\varrho u\right) - u \partial_x \left( \varrho u^2 + \mathcal{P}\right) +\mathfrak{f} \varrho u.
\end{equation}
Substituting this into the energy conservation law we obtain the final set of equations
\begin{align}\label{eqSM:continuityfull}
	&\partial_t \varrho  + \partial_x (\varrho u ) = 0, \qquad \partial_t (\varrho u)  + \partial_x \left( \varrho u^2 + \mathcal{P} \right) = \mathfrak{f} \varrho, \\
	&\partial_t \left( \varrho e \right) + \partial_x \left(u\varrho e + \mathcal{J}\right) +\mathcal{P} \partial_x u = 0,
\end{align}
reported in the main text. 

\section{Thermal boosted state}\label{secSM:sec3}

In this Section we study thermal boosted states. We show that a thermal boosted state characterized by chemical potentials $\beta_0$, $\beta_1$ and $\beta_2$ can be related to a thermal state with zero momentum and appropriately chosen $\bar{\beta}_0$, $\bar{\beta}_2$. Moreover, we will show that heat current $\mathcal{J}_v$ vanishes in such case and that the pressure $\mathcal{P}_v$ of a thermal boosted state is equal to the hydrostatic pressure $P$.

We start with the free energy density~\cite{lecture_notes_GHD}
\begin{equation}
\label{eqSM:freeene}
    \mathsf{f}= \frac{1}{2 \pi} \int \dd \lambda F[\epsilon(\lambda)],
\end{equation}
where function $F[\epsilon(\lambda)]$ depends on the particle' statistics. For example, for fermionic case relevant for Lieb-Liniger model~\cite{Yang1969} $F[\epsilon(\lambda)]=-\log(1+e^{-\epsilon})$. The pseudoenergy $\epsilon(\lambda)$ is determined from equation
\begin{equation}
\label{eqSM:YYeq}
    \epsilon(\lambda) =\epsilon_0(\lambda)+ \int \dd \lambda' \mathcal{T}(\lambda'-\lambda) F[\epsilon(\lambda')],
\end{equation}
where for thermal boosted state the bare pseudoenergy reads
\begin{equation}
    \epsilon_0(\lambda) =\beta_0 \mathscr{h}_0(\lambda) + \beta_1 \mathscr{h}_1(\lambda) +\beta_2 \mathscr{h}_2(\lambda).
\end{equation}
In a thermal state, on the other hand $\beta_0=-\mu/T, \, \beta_1=0, \, \beta_2= 1/T$ with $\mu,T$ equal to chemical potential and temperature, respectively (we set $k_B=1$).

We move now to discuss properties of a thermal boosted state, let $\rho_{\rm p}(\lambda)$ be its distribution function.
Then its first moment is non-zero,
\begin{equation}
	u = \frac{\int {\rm d}\lambda\, \lambda \rho_{\rm p}(\lambda)}{\int {\rm d}\lambda\, \rho_{\rm p}(\lambda)}.
\end{equation}
Define $\bar{\rho}_{\rm p}(\lambda) = \rho_{\rm p}(\lambda + u)$. We will show that $\bar{\rho}_{\rm p}(\lambda)$ is a thermal state and find expressions for $\bar{\beta}_0$ and $\bar{\beta}_2$ in terms of the chemical potentials of the boosted state. This can be argued on the basis of the Galilean invariance of the theory or computed explicitly. We follow the latter path.

Consider the total density $\rho_{\rm tot}(\lambda)$, it obeys the integral relation \eqref{eqSM:rhotot}.
Using the definition of $\bar{\rho}_{\rm p}(\lambda)$ it is easy to show that the corresponding total density is given by $\bar{\rho}_{\rm tot}(\lambda) = \rho_{\rm tot}(\lambda + u)$. From this follows that the  occupation functions and dressed pseudo-energies $\epsilon(\lambda)$ and $\bar{\epsilon}(\lambda)$ obey the same relation $\bar{\epsilon}(\lambda) = \epsilon(\lambda + u)$. Inspecting equation \eqref{eqSM:YYeq} we finally find that also $\bar{\epsilon}_0(\lambda) = \epsilon_0(\lambda + u)$. Writing this relation explicitly we have
\begin{equation}
	\bar{\beta}_0 + \frac{1}{2}\bar{\beta}_2 \lambda^2 = \beta_0 + \beta_1 (\lambda + u)  + \frac{1}{2}\beta_2 (\lambda + u)^2.
\end{equation}
Matching the terms with the same power of $\lambda$ we find
 \begin{equation}
 	\bar{\beta}_0 = \beta_0 + u \beta_1 + \frac{1}{2}u^2 \beta_2, \quad \bar{\beta}_2 = \beta_2, \quad \beta_1 +  u \beta_2 = 0.
 \end{equation}
The third equality is always fulfilled and gives $u = - \beta_1/\beta_2$. 

The last property follows from the observation that for $\lambda = u$ function $\rho_{\rm p}(\lambda)$ has the maximum. On the other hand
\begin{equation}
	\frac{\partial \rho_{\rm p}(\lambda)}{\partial \lambda} \sim \frac{\partial \epsilon(\lambda)}{\partial \lambda}.
\end{equation}
Function $\epsilon'(\lambda)$ obeys a linear integral equation
\begin{equation}
\label{eqSM:epsilonD}
	\epsilon'(\lambda) = \epsilon_0'(\lambda) + \int {\rm d}\mu \mathcal{T}(\lambda - \mu) n(\mu) \epsilon'(\mu).
\end{equation}
which follows from \eqref{eqSM:YYeq} and the identity~\cite{lecture_notes_GHD}
\begin{equation}
\label{eqSM:nFe}
    n=\frac{{\rm d} F}{{\rm d} \epsilon}
\end{equation}
and because of its linearity it maps zero to zero hence we need to solve $\epsilon_0'(\lambda) = 0$. Using $\epsilon_0(\lambda) = \beta_0 + \beta_1 \lambda + \frac{1}{2}\beta_2 \lambda^2$ we readily find the relation between $u$ and the chemical potentials.

We can also show using \eqref{eqSM:vinteq} that the effective velocity transforms as $\bar{v}(\lambda) + u = v(\lambda + u)$. This has a consequence that $v(u) = u$ where we used that $\bar{v}(0) = 0$. 

With these expressions we can show that the hydrodynamic pressure of a thermal boosted state reduces to the hydrostatic pressure $P$, see~\cite{Bouchoule2022} and the end of the present section. Hence the bulk viscosity vanishes is such case. We have
\begin{equation}
\label{eqSM:Phydro}
	\mathcal{P}_v = \int {\rm d}\xi\, \xi v(\xi + u) \rho_{\rm p}(\xi + u) = \int {\rm d}\xi\, \xi (\bar{v}(\xi) + u)\bar{\rho}_{\rm p}(\xi) = \int {\rm d}\xi\, \xi \bar{v}(\xi)\bar{\rho}_{\rm p}(\xi) = P.
\end{equation}
For the heat current we have
\begin{equation}
	\mathcal{J}_v= \frac{1}{2} \int {\rm d}\xi \xi^2 (v(u+\xi)-u) \rho_{\rm p}(u+\xi) = \frac{1}{2} \int {\rm d}\xi \xi^2 \bar{v}(\xi) \bar{\rho}_p(\xi) = 0,
\end{equation}
because $\bar{v}(\xi)$ is an odd function while $\bar{\rho}_p(\xi)$ is even. 

The entropy of the boosted state is equal to the entropy of the unboosted state. Recall the formula for the entropy density
\begin{equation}
	\mathsf{s} = \int {\rm d}\lambda \rho_{\rm tot}(\lambda) g(n(\lambda)),
\end{equation}
where function $g(n)$ depends on particles' statistics~\cite{lecture_notes_GHD, hydro_diff_prl}. For the fermionic models we have $g[n] = n \log n + (1-n) \log (1-n)$. In general $ g=\epsilon(\lambda)n(\lambda)-F[\epsilon(\lambda)]$.
Using now the relations between the particle distribution and filling function of the boosted and unboosted state we see that the entropies are indeed equal.

\subsection*{Expression for the hydrostatic pressure in terms of $v(\lambda)$}
Let us close this section by further validating equation \eqref{eqSM:Phydro}. Our aim is to show that indeed
\begin{equation}
    P = \int \dd \lambda   \lambda  v(\lambda)\rho_{\rm p}(\lambda)
\end{equation}
where $\rho_{\rm p}(\lambda)$ is thermal state, is the hydrostatic pressure as defined in thermodynamics, see eq.~\eqref{eqSM:thermalf} of the next section. To this end we  generalize the computations for the Lieb-Liniger model presented in~\cite{Bouchoule2022}.

We start by noting that $\rho_{\rm tot}(\lambda)= \frac{1}{2 \pi} \mathscr{h}_0^{\rm dr}(\lambda)$ and $v(\lambda)=\mathscr{h}_1^{\rm dr}(\lambda)/\mathscr{h}_0^{\rm dr}(\lambda)$, hence
\begin{equation}
\label{eqSM:Pj1}
    P = \frac{1}{2 \pi} \int \dd \lambda\,  \lambda \,  n(\lambda) \mathscr{h}_1^{\rm dr}(\lambda).
\end{equation}
Note that we have also shown at this point that $P=\mathscr{j}_1$, where $\mathscr{j}_1$ is the momentum current  (compare with formula for current expectation value in~\cite{lecture_notes_GHD}). To get $\mathscr{h}_1^{\rm dr}(\lambda)$ we look at \eqref{eqSM:epsilonD}. For thermal state it reads
\begin{equation}
    \epsilon'(\lambda)= \frac{1}{T}\mathscr{h}_1(\lambda) + \int \dd \lambda' \mathcal{T}(\lambda-\lambda') n(\lambda') \epsilon'(\lambda').
\end{equation}
Comparing with the definition of dressing operation \eqref{eqSM:dressing} we conclude that $\mathscr{h}_1^{\rm dr}(\lambda)=T \epsilon ' (\lambda)$. Moreover, using \eqref{eqSM:nFe} and integrating by parts we get
\begin{equation}
    P= -\frac{T}{2 \pi} \int \dd \lambda F[\epsilon(\lambda)] = -T \mathsf{f}
\end{equation}
in agreement with \eqref{eqSM:thermalf}.

\section{Thermodynamics and hydrodynamic matrices}\label{secSM:sec4}

In this section we will relate the matrix elements of hydrodynamic matrices~\cite{DrudeWeight_LL} $A$, $B$ and $C$ to the thermodynamics of the integrable model. This will be very helpful in linking results from linearized GHD with thermodynamic quantities, which appear naturally in the Navier-Stokes equations.

We start by defining the free energy flux~\cite{lecture_notes_GHD}, in addition to the free energy density $\mathsf{f}$ defined in~\eqref{eqSM:freeene}, 
\begin{equation}
    \mathsf{g}= \frac{1}{2 \pi} \int \dd \lambda E'(
    \lambda) F[\epsilon(\lambda)],
\end{equation}
where the pseudoenergy $\epsilon(\lambda)$ satisfies \eqref{eqSM:YYeq}. 

For a generic GGE state, the bare pseudoenergy includes higher conserved charges~\cite{Mossel2012}. The simplest choice being the ultra-local charges $\mathscr{q}_j$, 
\begin{equation}
	\epsilon_0(\lambda) = \sum_{j=0}^{\infty} \beta_j \mathscr{h}_j(\lambda).
\end{equation}
The expectation values of the charges in the GGE state follow then from derivatives of the free energy density with respect to the generalized temperatures. In a similar fashion, the expectation values of the currents are given by the derivatives of the free energy flux,
\begin{equation}
	\mathscr{q}_i = \frac{\partial \mathsf{f}}{\partial \beta_i}, \qquad \mathscr{j}_i = \frac{\partial \mathsf{g}}{\partial \beta_i}.
\end{equation}
The hydrodynamic matrices $\mathcal{B}$ and $\mathcal{C}$ are susceptibilities of the currents and charges
\begin{equation}
	\mathcal{B}_{i,j} = - \frac{\partial \mathscr{j}_i}{\partial \beta_j} = \int {\rm d}\lambda\, \rho_{\rm p}(\lambda) f(\lambda) v(\lambda) \mathscr{h}_i^{\rm dr}(\lambda) \mathscr{h}_j^{\rm dr}(\lambda), \qquad \mathcal{C}_{i,j} =-\frac{\partial \mathscr{q}_i}{\partial \beta_j} =\int {\rm d} \lambda\, \rho_{\rm p}(\lambda) f(\lambda)  \mathscr{h}_i^{\rm dr}(\lambda) \mathscr{h}_j^{\rm dr}(\lambda).
\end{equation}
The matrices defined above are clearly symmetric. Above we defined 
\begin{equation}
\label{eqSM:f_factor}
    f(\lambda)=- \frac{{\rm d}^2F/{\rm d} \epsilon^2}{{\rm d}F/{\rm d} \epsilon},
\end{equation}
which is called the statistical factor~\cite{lecture_notes_GHD}. For instance, in the case of fermionic statistics one finds $f(\theta)=1-n(\theta)$.
We also note that for Galilean invariant systems there is an useful relation $\mathcal{B}_{1,0}= \mathcal{C}_{1,1}$. 

The hydrodynamic matrices depend on the choice of the basis for the ultra-local charges. For the thermodynamics it is natural to work with matrix elements of the ultralocal charges $\mathscr{h}_n(\lambda)$ as defined above. Instead, for the hydrodynamics, it is more convenient to work with the orthonormalized basis obtained from the ultra-local charges through the Gram-Schmidt method with the inner product~\eqref{eqSM:Innerprod}. For $h_n(\lambda)$ denoting the single-particle eigenvalues of the resulting charges, the hydrodynamic matrices are
\begin{equation}
	B_{i,j} = - \frac{\partial j_i}{\partial \beta_j} = \int {\rm d}\lambda\, \rho_{\rm p}(\lambda) f(\lambda) v(\lambda) h_i^{\rm dr}(\lambda) h_j^{\rm dr}(\lambda) , \qquad C_{i,j} =-\frac{\partial q_i}{\partial \beta_j} =\int {\rm d} \lambda\, \rho_{\rm p}(\lambda) f(\lambda) h_i^{\rm dr}(\lambda) h_j^{\rm dr}(\lambda)=\delta_{i,j}.
\end{equation}

In the remainder of this section we will relate matrix elements of $\mathcal{B}$ and $\mathcal{C}$ to thermodynamic quantities. We focus on the Gibbs state with a thermal $\epsilon_0(\lambda)$ function,
\begin{equation}
    \epsilon_0(\lambda) = -\frac{\mu}{T} \mathscr{h}_0(\lambda) + \frac{1}{T} \mathscr{h}_2(\lambda).
\end{equation}
We also introduce the notation
\begin{equation}
    \varrho=\mathscr{q}_0= \int \dd \lambda \mathscr{h}_0 (\lambda) \rho_{\rm p}(\lambda), \qquad \mathsf{e} =\mathscr{q}_2= \int \dd \lambda \mathscr{h}_2(\lambda) \rho_{\rm p} (\lambda), \qquad \mathsf{s}= \int \dd \lambda \rho_{\rm tot}(\lambda) g[n(\lambda)],  
\end{equation}
for particle, energy and entropy density, respectively. Moreover, by 
\begin{equation}
    s=\varrho^{-1} \mathsf{s}, \qquad e= \varrho^{-1} \mathsf{e},
\end{equation}we denote entropy and energy per particle. For Gibbs state the free energy $\mathsf{f}$ takes the form of the grand potential per unit length divided by temperature and can be written as
\begin{equation}
\label{eqSM:thermalf}
    \mathsf{f} = - \frac{\mu}{T} \varrho +\frac{1}{T} \mathsf{e} - \mathsf{s}=-\frac{P}{T}.
\end{equation}
With this, we compute various thermodynamic quantities. We start with the sound velocity.
\subsection{Sound velocity}
The sound velocity is related to the adiabatic compressibility $\varkappa_S$ and defined as (we work with unit mass)
\begin{equation}
\label{eqSM:sounddef}
    v_s^2 = \Partial{P}{\varrho}{s} = \frac{1}{ \varrho  \varkappa_S}, \qquad \varkappa_S =\frac{1}{\varrho} \Partial{\varrho}{P}{s}.
\end{equation}
To compute the relevant derivative, we realize that $P,\varrho$ and $s$ depend only on two thermodynamic parameters, which specify the state: $\mu,T$. Equivalently, they depend on $\beta_0,\beta_2$. We may thus write
\begin{equation}
    \dd s= \Partial{s}{\beta_0}{\beta_2} \dd \beta_0+ \Partial{s}{\beta_2}{\beta_0} \dd \beta_2,
\end{equation}
and similarly for $\dd \varrho$ and $\dd P$. To keep the entropy fixed $\dd \beta_0$ and $\dd \beta_2$ must be related through
\begin{equation}
    \dd \beta_0 = -\frac{\Partial{s}{\beta_2}{\beta_0}}{\Partial{s}{\beta_0}{\beta_2}} \dd \beta_2,
\end{equation}
which leads to the following formula 
\begin{equation} \label{eqSM:adiabatic_compressibility_partial}
    \Partial{\varrho}{P}{s} = \frac{\Partial{\varrho}{\beta_0}{\beta_2}\Partial{s}{\beta_2}{\beta_0}-\Partial{\varrho}{\beta_2}{\beta_0}\Partial{s}{\beta_0}{\beta_2}}{\Partial{P}{\beta_0}{\beta_2}\Partial{s}{\beta_2}{\beta_0}-\Partial{P}{\beta_2}{\beta_0}\Partial{s}{\beta_0}{\beta_2}}.
\end{equation}
The derivatives of the entropy can be computed from the expression for free energy \eqref{eqSM:thermalf}. The entropy per particle equals to
\begin{equation}
    s = \frac{\beta_0 \varrho + \beta_2 \mathsf{e} -\mathsf{f}}{\varrho},
\end{equation}
and the two required derivatives are
\begin{equation}
    \Partial{s}{\beta_0}{\beta_2}=\frac{\mathcal{C}_{0,0}(\beta_2 \mathsf{e} -\mathsf{f})- \mathcal{C}_{2,0} \beta_2 \varrho}{\varrho^2}, \qquad 
    \Partial{s}{\beta_2}{\beta_0}=\frac{\mathcal{C}_{2,0}(\beta_2 \mathsf{e} -\mathsf{f})- \mathcal{C}_{2,2} \beta_2 \varrho}{\varrho^2}.
\end{equation}
We use now the relation $\mathscr{j}_1=P$, valid for Galilean-invariant models, see \eqref{eqSM:Pj1}. Together with $P=-T\mathsf{f}=-\mathsf{f}/\beta_2$ they yield
\begin{equation}
    \mathcal{B}_{1,0}=-\Partial{P}{\beta_0}{\beta_2}=\frac{\varrho}{\beta_2}=T \varrho, \qquad 
    \mathcal{B}_{2,1}=-\Partial{P}{\beta_2}{\beta_0}=\frac{\beta_2 \mathsf{e}-\mathsf{f}}{\beta_2^2},
\end{equation}
which gives
\begin{equation}
    \Partial{s}{\beta_0}{\beta_2}=\frac{1}{\mathcal{B}_{1,0}^2}\left(\mathcal{C}_{0,0}\mathcal{B}_{2,1}- \mathcal{C}_{2,0}\mathcal{B}_{1,0} \right), \qquad     \Partial{s}{\beta_2}{\beta_0}=\frac{1}{\mathcal{B}_{1,0}^2}\left(\mathcal{C}_{2,0}\mathcal{B}_{2,1}- \mathcal{C}_{2,2}\mathcal{B}_{1,0} \right).
\end{equation}
We now go back to~\eqref{eqSM:adiabatic_compressibility_partial} and~\eqref{eqSM:sounddef} to find the final expression for the adiabatic compressibilty
\begin{equation} \label{eqSM:k_S}
    \varkappa_S =T \frac{\mathcal{C}_{2,2}\mathcal{C}_{0,0}-\mathcal{C}_{2,0}^2}{\mathcal{B}_{1,0}^2 \mathcal{C}_{2,2}+\mathcal{B}_{2,1}^2\mathcal{C}_{0,0}-2 \mathcal{B}_{1,0}\mathcal{B}_{2,1}\mathcal{C}_{2,0}},
\end{equation}
from which the final expression for the sound velocity follows
\begin{equation}
\label{eqSM:soundvelocity}
   v_s = \sqrt{\frac{1}{\mathcal{B}_{1,0}} \frac{\mathcal{B}_{1,0}^2 \mathcal{C}_{2,2}+\mathcal{B}_{2,1}^2\mathcal{C}_{0,0}-2 \mathcal{B}_{1,0}\mathcal{B}_{2,1}\mathcal{C}_{2,0}}{\mathcal{C}_{2,2}\mathcal{C}_{0,0}-\mathcal{C}_{2,0}^2}}.
\end{equation}
We have analytically confirmed that this general formula reduces to known formulas for free fermions $v_s^{\rm Fermi}=\sqrt{3T \frac{f_{3/2}(e^{\mu/T})}{f_{1/2}(e^{\mu/T})}}$ where $f_\alpha(x)$ are Fermi functions and for hard-rods~\cite{Dorfman2021} $v_s^{\rm HR}= \sqrt{3T/(1-a \varrho)^2}$, where $a $ is rod's length. For the Lieb-Liniger model we confirmed with derivatives computed numerically from TBA.
\subsection{Specific heats, isothermal compressibility and additional identities}
We calculate in a similar fashion several more thermodynamic quantities, which are useful for purposes of this work.
{\bf Isothermal compressibility:} The isothermal compressibility is defined as 
\begin{equation}
    \varkappa_T= \frac{1}{\varrho} \Partial{\varrho}{P}{T},
\end{equation}
and standard manipulations give
\begin{equation} \label{eqSM:k_T}
	\varkappa_T= \frac{1}{\varrho} \frac{\Partial{\varrho}{\beta_0}{\beta_2}}{\Partial{P}{\beta_0}{\beta_2}}= \frac{1}{\varrho} \frac{\mathcal{C}_{0,0}}{\mathcal{B}_{1,0}}.
\end{equation}

{\bf Specific heat at constant volume:} The definition is
\begin{equation}
    c_V=\Partial{e}{T}{\varrho}=\frac{1}{\varrho}\Partial{\mathsf{e}}{T}{\varrho}.
\end{equation}
Expressing $\dd T$ with $ \dd \beta_0$ and $\dd \beta_2$ we obtain 
\begin{equation}
    c_V=\frac{1}{\varrho} \frac{-\Partial{\mathsf{e}}{\beta_0}{\beta_2}\Partial{\varrho}{\beta_2}{\beta_0}+\Partial{\mathsf{e}}{\beta_2}{\beta_0}\Partial{\varrho}{\beta_0}{\beta_2}}{-T^2 \Partial{\varrho}{\beta_0}{\beta_2}},
\end{equation}
which can now be expressed with the hydrodynamic matrices as
\begin{equation} \label{eqSM:c_V}
    c_V=\frac{1}{T}\frac{\mathcal{C}_{2,2}\mathcal{C}_{0,0}-\mathcal{C}_{2,0}^2}{\mathcal{C}_{0,0} \mathcal{B}_{1,0}}.
\end{equation}

{\bf Specific heat at constant pressure:} The definition is
\begin{equation}
    c_P=T \Partial{s}{T}{P}.
\end{equation}
Proceeding in an analogous way as in the previous cases we find
\begin{equation} \label{eqSM:c_P}
    c_P =  \frac{1}{T} \frac{\mathcal{B}_{1,0}^2 \mathcal{C}_{2,2}+\mathcal{B}_{2,1}^2\mathcal{C}_{0,0}-2 \mathcal{B}_{1,0}\mathcal{B}_{2,1}\mathcal{C}_{2,0}}{\mathcal{B}_{1,0}^3}.
\end{equation}
As a check for our expressions, we also observe that the two compressibilities and two specific heats are not independent from each other. Instead, they obey the thermodynamic identity
\begin{equation}\label{eqSM:cpcvkappa}
    \varkappa_S = \frac{c_V}{c_P} \varkappa_T.
\end{equation}
One can easily verify that this identity holds for the expressions given in~\eqref{eqSM:k_S},~\eqref{eqSM:k_T},~\eqref{eqSM:c_V} and~\eqref{eqSM:c_P}.

{\bf Expressions for the matrix elements of $A$:}
The sound velocity may be rewritten with $A$ matrix elements in GS-orthonormalized basis. These matrix elements are defined in \eqref{eqSM:matrixelements} and appear explicitly in the equation \eqref{eqSM:transporteigenproblem}, from which dispersion relations are determined. It is thus useful to write thermodynamic quantities in terms of $A$ matrix in this basis
\begin{equation}\label{eqSM:thermoAmatrix}
    A_{1,0}=\sqrt{\frac{\mathcal{B}_{1,0}}{\mathcal{C}_{0,0}}}= \frac{1}{\sqrt{\varrho \varkappa_T}}, \qquad 
    A_{2,1}= \frac{\mathcal{B}_{2,1}\mathcal{C}_{0,0}-\mathcal{B}_{1,0}\mathcal{C}_{2,0}}{\sqrt{\mathcal{C}_{0,0} \mathcal{B}_{1,0} (\mathcal{C}_{2,2} \mathcal{C}_{0,0}- \mathcal{C}_{2,0}^2)}}=\frac{T}{\varrho^2 c_V} \Partial{P}{T}{\varrho}= \frac{1}{\sqrt{\varrho \varkappa_T}} \sqrt{c_P/c_V-1}.
\end{equation}
Above we have used a general thermodynamic identity~\cite{ResiboisBOOK}
\begin{equation}\label{eqSM:cp-cv}
    c_P-c_V=\frac{T}{\varrho^2}\frac{\Partial{P}{T}{\varrho}^2}{\Partial{P}{\varrho}{T}}.
\end{equation}
In the perturbation theory, we will find that speed of sound becomes 
\begin{equation}
    v_s=\sqrt{A_{1,0}^2+A_{2,1}^2},
\end{equation}
which agrees with \eqref{eqSM:soundvelocity}. Moreover, we observe that
\begin{equation}
    \frac{A_{2,1}^2}{A_{1,0}^2}=c_P \left( \frac{1}{c_V}-\frac{1}{c_P} \right)= \frac{c_P}{c_V}-1.
\end{equation}
{\bf Additional identities:}
For analysis of the structure of heat and sound modes the following identities are useful
\begin{equation}\label{eqSM:additionalthermo}
    \Partial{P}{T}{\varrho}= \frac{1}{T^2} \frac{\mathcal{B}_{2,1} \mathcal{C}_{0,0}- \mathcal{B}_{1,0} \mathcal{C}_{2,0}}{\mathcal{C}_{0,0}},  \qquad
    \frac{1}{\varrho} \left(\mathsf{e}+P - T \Partial{P}{T}{\varrho}\right) = \frac{\mathcal{C}_{2,0}}{\mathcal{C}_{0,0}}.
\end{equation}
They can be derived with the methods presented above. Lastly, let us comment that the method presented in this section gives thermodynamic quantities as explicit functionals (that is, combinations of matrix elements of hydrodynamic matrices) of the thermal rapidity distribution $\rho_{\rm p}(\lambda)$. This representation is convenient and may be useful for applications which go beyond the scope of this work.

\section{Derivation of the linearized equation for charge perturbations}\label{secSM:sec5}

We again invoke the GHD equation supplemented with the collision integral term, this time with $U(x)=0$,
\begin{equation}
\label{SM:GHD2}
	\partial_t \rho_{\rm p} + \partial_x \left(v_{\rho} \rho_{\rm p} \right) = \frac{1}{2} \partial_x \left(\mathfrak{D}_{\rho}\partial_x \rho_{\rm p} \right) + \mathcal{I}[\rho_{\rm p}].
\end{equation}
Our aim here is to first linearize this equation and then rewrite it as a continuity equation for the densities of local charges. 
To this end we linearize it around a stationary uniform thermal state $\rho(x,t) = \rho_{\rm th}+ \delta \rho_{\rm p}(x,t)$
getting equation
\begin{equation} \label{SM:GHD3}
    \partial_t \delta \rho_{\rm p}+\partial_x \left((1-n \mathcal{T})^{-1}v_{\rho}(1-n \mathcal{T}) \delta \rho_{\rm p} \right) = \frac{1}{2} \mathfrak{D}_{\rm th} \partial_x^2 \delta \rho_{\rm p} + \left(\frac{\delta \mathcal{I}}{\delta \rho_{\rm p}}\right)_{\rm th}\delta \rho_{\rm p}.
\end{equation}
The second term in the equation above (linearized expression for effective velocity term) was derived in~\cite{DrudeWeight_LL}.

Motivated by the structure of $\Gamma$ matrix that we consider (see Sec.~\ref{secSM:sec9}), we wish to change the degrees of freedom from a perturbation of quasiparticle distribution to a perturbation of the bare pseudoenergy $\epsilon_0$. To establish a relation between them, we start with 
\begin{equation}
    \rho_{\rm p}= n \rho_{\rm tot}, \qquad \delta \rho_{\rm p}= \delta n \rho_{\rm tot}+ n \delta \rho_{\rm tot}.
\end{equation}
The perturbation occupation function may be related to the perturbation of pseudoenergy via \eqref{eqSM:nFe}, hence $\delta n = - f n \delta \epsilon$. Moreover from \eqref{eqSM:rhotot} we have $\delta \rho_{\rm tot} =\mathcal{T} \delta \rho_{\rm p}$ and from \eqref{eqSM:YYeq} we get $\delta \epsilon = (1-\mathcal{T}n)^{-1} \delta \epsilon_0$. Combining, we find the following relation
\begin{equation}
\label{eqSM:deltaepsrelation}
    \delta \rho_{\rm p} = -(1-n \mathcal{T})^{-1} \rho_{\rm p} f \delta \epsilon= -(1-n \mathcal{T})^{-1} \rho_{\rm p} f (1-\mathcal{T}n )^{-1}\delta \epsilon_0 = - \mathcal{C} \delta \epsilon_0.
\end{equation}
where the charge susceptibility $\mathcal{C}$ operator was introduced in \eqref{eqSM:Ckernel} and the operator $(1-\mathcal{T}n)^{-1}$ implements the dressing \eqref{eqSM:dressing}. Function $\delta\epsilon_0(\lambda)$ can be expanded as 
\begin{equation}
	\delta \epsilon_0(\lambda) = \sum_{m} ( h_m|\delta \epsilon_0) h_m(\lambda),
\end{equation}
where $h_m(\lambda)$ form an orthonormal set of functions with respect to the hydrodynamic scalar product $(\cdot|\cdot)$, see also~\cite{lecture_notes_GHD},
\begin{equation}
\label{eqSM:Innerprod}
    (h|g) = \int {\rm d} \lambda\, \rho_{\rm p}(\lambda) f(\lambda) h^{\rm dr}(\lambda)g^{\rm dr}(\lambda).
\end{equation}
In practice, $h_m$ are constructed by the Gram-Schmidt orthonormalization of the ultra-local charges with single particle eigenvalue $\mathscr{h}_m(\lambda) = \lambda^m/m!$. The most important first three functions are
\begin{equation}
\label{eqSM:charges012}
    h_0(\lambda)=\frac{1}{\sqrt{\mathcal{C}_{0,0}}} \mathscr{h}_0(\lambda), \qquad h_1(\lambda)=\frac{1}{\sqrt{\mathcal{C}_{1,1}}} \mathscr{h}_1(\lambda), \qquad h_2(\lambda)=\sqrt{\frac{\mathcal{C}_{0,0}}{\mathcal{C}_{2,2}\mathcal{C}_{0,0}-\mathcal{C}_{2,0}^2}} \left( \mathscr{h}_2(\lambda) - \frac{\mathcal{C}_{2,0}}{\mathcal{C}_{0,0}}  \mathscr{h}_0(\lambda) \right ).
\end{equation}
The expansion coefficients are directly related to the change in the expectation value of the $m$-th conserved charge, $( h_m|\delta \epsilon_0) = - \delta q_m$, where
\begin{equation} \label{SM:change_charge}
	\delta q_m = \int {\rm d}\lambda\, h_m(\lambda) \delta\rho_{\rm p}(\lambda).
\end{equation}
This relation follows from the following identity~\cite{lecture_notes_GHD} (here dot-product is $h \cdot g=\int {\rm d} \lambda h(\lambda) g(\lambda)$)
\begin{equation}
\label{eqSM:symmid}
     (1-n \mathcal{T})^{-1}n g \cdot h=g \cdot n(1-\mathcal{T}n)^{-1}h = g \cdot n h^{\rm dr},
\end{equation}
which follows from the symmetry of the kernel $\mathcal{T}$. With the help of this identity we find 
\begin{align}
     \delta q_m &=-\int {\rm d} \lambda\, h_m(\lambda)(1-n \mathcal{T})^{-1} \rho_{\rm p} f(1-\mathcal{T}n )^{-1} \delta \epsilon_0 \nonumber \\
     &= -\int {\rm d}\lambda\, \rho_{\rm p}(\lambda) f(\lambda) h_m^{\rm dr}(\lambda)\delta \epsilon_0^{\rm dr}(\lambda)= - (h_m|\delta \epsilon_0),\label{eqSM:inneprod}
\end{align}
where we used in the second step with $h=h_m$ and $g=\rho_{\rm p} f \delta \epsilon_0^{\rm dr}/n$. This allows us to write
\begin{equation}
\label{eqSM:delrgorep}
	\delta \rho_{\rm p} = (1-n \mathcal{T})^{-1} \rho_{\rm p} f (1-\mathcal{T}n)^{-1} \sum_{m}  h_m \delta q_m.
\end{equation}
Functions $h_n$ allow us to reformulate the GHD equations as equations for the charge perturbations. We multiply the GHD equation~\eqref{SM:GHD3} by $h_n(\lambda)$, plug \eqref{eqSM:delrgorep} and integrate over $\lambda$ taking advantage of the orthonormality of functions $h_n$. Moreover, we introduce $\Gamma$ operator 
\begin{equation}
    \Gamma = \frac{\delta \mathcal{I}}{\delta \epsilon_0}= - \frac{\delta \mathcal{I}}{\delta \rho_{\rm p}} \mathcal{C}. 
\end{equation}
The result is 
\begin{equation}
\label{SM:linear}
    \partial _t \delta q_n + \sum_m A_{nm} \partial_x \delta q_m = \sum_m D_{nm} \partial_x^2 \delta q_m -\sum_m \Gamma_{nm} \delta q_m,
\end{equation}
with
\begin{align}\label{eqSM:matrixelements}
    A_{nm} &=B_{nm}=  \int {\rm d} \lambda\, \rho_{\rm p}(\lambda)f(\lambda) v(\lambda) h_n^{\rm dr}(\lambda) h_m^{\rm dr}(\lambda),\\
    D_{nm} &= \frac{1}{2}\int {\rm d}\lambda\, {\rm d} \mu\,  h_n^{\rm dr}(\lambda) \rho_{\rm tot}^{-1}(\lambda) \tilde{\mathfrak{D}}_{\rm th}(\lambda, \mu)\rho_{\rm tot}^{-1}(\mu) \rho_{\rm p}(\mu)f(\mu) h_m^{\rm dr}(\mu),\\
    \Gamma_{nm} &= \int {\rm d}\lambda\, {\rm d} \mu\, h_n(\lambda) \Gamma(\lambda, \mu) h_m(\mu).
\end{align}

Furthermore, we can take into account that $\Gamma$ operator hides the Dressing operation~\cite{weak_integ_breaking,PGK, PhysRevLett.124.140603}. The Dressing operation is defined through (where $F(\mu|\lambda)$ is backflow function),
\begin{equation}
    f^{\rm Dr}(\lambda)= f(\lambda) - \int {\rm d} \mu\, n (\mu) F(\mu|\lambda) f'(\mu),
\end{equation}
such that the collision integral is (note that operator $\mathbb{F}$ effectively implements Dressing)
\begin{equation}
\label{eqSM:Foperator}
    \Gamma(\lambda,\mu) =\int {\rm d} \nu\, {\rm d} \nu' \,\mathbb{F}(\lambda,\nu) \Gamma_0(\nu, \nu')\mathbb{F}(\mu,\nu'), \qquad \mathbb{F}(\lambda,\nu)= \delta(\lambda-\nu)+ \partial_\lambda(n(\lambda) F(\lambda|\nu)) ,
\end{equation}
where $\Gamma_0$ is the kernel of linearized bare  collision integral. This Dressing operation may be shifted to $h_n(\lambda)$ so finally we get
\begin{equation}
    \Gamma_{nm}=\int {\rm d}\lambda\, {\rm d} \mu\, h_n^{\rm Dr}(\lambda) \Gamma_0(\lambda, \mu) h_m^{\rm Dr}(\mu).
\end{equation}

Let us comment on general properties of the matrix elements listed above. Matrix elements $A_{nm}$ are non-zero where $n,m$ are of different parity, instead $\Gamma_{nm}$ and $D_{nm}$ are non-zero only when $n,m$ have the same parity. We also have $\Gamma_{nm} = 0$ for $n,m=0,1,2$, which reflects the $3$ collision invariants, whereas $D_{0,n}=D_{n,0} = 0$ which follows from the Markovianity of the diffusion kernel discussed in Section~\ref{secSM:sec1}.

For the spectrum of $\Gamma$ operator we assume that it is non-negative and gapped. The eigenvectors correspond to either even or odd functions of rapidities. We confirm these properties for the collision integral analyzed in Sec.~\ref{secSM:sec9}.

\section{Linearized Navier-Stokes equations and transport coefficients}\label{secSM:sec6}
We linearize the Navier-Stokes equations, which can be understood as \eqref{eqSM:continuityfull} with the phenomenological expressions
\begin{align} \label{eqSM:pheno}
	\mathcal{P} = P - \zeta \partial_x u, \qquad \mathcal{J} = - \kappa \partial_x T,
\end{align}
for the dynamic pressure and heat current. In this section, we do not specify $\zeta$ and $\kappa$. Our aim is to focus on general structure of N-S equations and find relation between the spectrum of the hydrodynamic modes and the transport coefficients.

We start by introducing the temperature field to the Navier-Stokes equations. For this, we note the following general thermodynamic equality, which can be derived from basic principles~\cite{ResiboisBOOK}
\begin{equation}
\label{eqSM:thermID}
	\delta T = \frac{1}{c_V} \delta e - \frac{1}{c_V}\left(\frac{P}{\varrho^2}-\frac{T}{\varrho^2} \Partial{P}{T}{\varrho}\right) \delta \varrho.
\end{equation}
One-dimensional version of Navier-Stokes equations~\cite{Balescu1975,Chakraborti2021} for the fields $\varrho(x,t)$, $u(x,t)$ and $T(x,t)$ reads then
\begin{equation}
\begin{aligned}
\label{eqSM:NS1}
    \partial_t \varrho &=-\partial_x (\varrho u), \\
    \partial_t u &= -u \partial_x u- \frac{1}{\varrho} \Partial{P}{\varrho}{T} \partial_x \varrho - \frac{1}{\varrho}\Partial{P}{T}{\varrho} \partial_x T+ \frac{\zeta}{\varrho}\partial_x^2 u,\\
    \partial_t T &= -\frac{T}{\varrho c_V} \Partial{P}{T}{\varrho}\partial_x u-u\partial_xT+\frac{\kappa}{\varrho c_V}\partial_x^2 T+\frac{\zeta}{\varrho c_V}(\partial_x u)^2,
    \end{aligned}
\end{equation}
where $\kappa, \zeta$ are thermal conductivity and bulk viscosity, respectively. Linearizing around a homogenous thermal state
\begin{equation}
    \varrho(x,t) = \varrho+ \delta \varrho (x,t), \qquad u(x,t) = 0+ \delta u(x,t), \qquad T(x,t) = T+ \delta T(x,t),
\end{equation}
one gets
\begin{equation}
	\begin{aligned}
   	\partial_t \delta \varrho &= - \varrho \partial_x \delta u,\\
	\partial_t \delta u &= - \frac{1}{\varrho} \Partial{P}{\varrho}{T} \partial_x \delta \varrho - \frac{1}{\varrho} \Partial{P}{T}{\varrho} \partial_x \delta T +\frac{\zeta}{\varrho} \partial_x^2 \delta u,\\
	\partial_t \delta T &= - \frac{T}{\varrho c_V} \Partial{P}{T}{\varrho} \partial_x \delta u+ \frac{ \kappa}{\varrho c_V}\partial_x^2 \delta T.
	\end{aligned}
\end{equation}
Moving to the $k$-space and anticipating the time-dependence as
\begin{equation}
    \delta \varrho (x,t) = \int \dd k  e^{ i kx - \Lambda t} \delta \varrho(k), \qquad \delta u(x,t) = \int \dd k  e^{ i kx - \Lambda t} \delta u(k), \qquad \delta T (x,t) = \int \dd k e^{ i kx - \Lambda t} \delta T(k),
\end{equation}
we get an eigenproblem with non-Hermitian matrix. As the matrix is non-Hermitean the left- and right-eigenvectors are different. This problem be cured by changing the variables as
\begin{equation}
    \delta \varrho = \sqrt{T \varrho /(\partial P/ \partial \varrho)_T} \, \delta \tilde{\varrho} = \varrho \sqrt{T \varkappa_T}\, \delta \tilde{\varrho}, \qquad \delta u =\sqrt{T/\varrho} \,\delta \tilde{u}, \qquad \delta T= T/\sqrt{\varrho c_V} \,  \delta \tilde{T}.
\end{equation}
In these variables, the relevant matrix becomes symmetric. Explicitly, when looking for eigenvalues $\Lambda$ we will deal with
\begin{equation}\label{eqSM:NSmatrix}
    \text{det}\begin{pmatrix}
    -\Lambda & i \frac{1}{\sqrt{\varrho \varkappa_T}} k & 0 \\
    i  \frac{1}{\sqrt{\varrho \varkappa_T}} k & -\Lambda+ \frac{\zeta}{\varrho} k^2 & i \sqrt{\frac{T}{\varrho^2 c_V}} \Partial{P}{T}{\varrho} k \\
    0 & i \sqrt{\frac{T}{\varrho^2 c_V}} \Partial{P}{T}{\varrho} k &-\Lambda +  \frac{\kappa}{\varrho c_V} k^2
    \end{pmatrix}=0.
\end{equation}
We use now the general thermodynamic identities \eqref{eqSM:cpcvkappa}, \eqref{eqSM:cp-cv} and $\frac{c_P}{c_V}\Partial{P}{\varrho}{T}=v_s^2$~\cite{ResiboisBOOK} and find the following dispersion relations
\begin{equation} \label{eqSM:SM_spectrum_NS}
\begin{aligned}
	 \Lambda_{\pm}(k) = \pm i v_s k +\frac{1}{2\varrho} \bigg[ \left( \frac{1}{c_V}-\frac{1}{c_P} \right) \kappa + \zeta \bigg] k^2 +O(k^3), \qquad 
    	\Lambda_{\rm th}(k) = \frac{ \kappa}{\varrho c_P}k^2 + O(k^3).
\end{aligned}
\end{equation}
These dispersion relations correspond to two sound modes and one heat mode, respectively. We observe that in the case of vanishing bulk viscosity $\zeta=0$, the ratio of the quadratic coefficients is $(c_P/c_V - 1)/2$ and therefore is universal. Depends only on the thermodynamics of the underlying gas but not on the integrability breaking terms. Vanishing bulk viscosity is an expected result for an ideal gas perturbed with weak interactions~\cite{Balescu1975}. We will see that in our setup the bulk viscosity vanishes in the Tonks-Girardeau ($c\rightarrow \infty$)  and in $c \to 0$ limits, see Sec.~\ref{secSM:sec8}. 

In addition to the eigenvalues, we can find eigenvectors. In the limit of $k \to 0$ we have
\begin{equation}\label{eqSM:NSheat&sound}
    \delta_{\pm} (k \to 0)= \frac{1}{\sqrt{2}} \left(\sqrt{\frac{c_V}{c_P}} \delta \tilde{\varrho} \pm \delta \tilde{u} +\sqrt{\frac{c_P-c_V}{c_P}} \delta \tilde{T} \right), \qquad \delta_{\rm th}(k \to 0) = - \sqrt{\frac{c_P-c_V}{c_P}} \delta \tilde{\varrho}+ \sqrt{\frac{c_V}{c_P}} \delta \tilde{T}.
\end{equation}
The general structure revealed in this section will be our reference point for perturbative treatment of \eqref{SM:linear}. It  will allow us to identify Navier-Stokes hydrodynamics appearing in GHD equation supplemented with a collision term. 

\section{Perturbation theory}\label{secSM:sec7}

As discussed in the main text the linearized problem \eqref{SM:linear} is solved by Fourier transform followed by perturbative solution in $k$. The dynamics of a single mode is
\begin{equation}
	\delta q_n(x,t) \sim e^{ikx - \Lambda(k) t} \delta q_n(k).
\end{equation}
Substituting this expression to the linearized GHD equations we obtain the following spectral problem
\begin{equation}\label{eqSM:transporteigenproblem}
	\left(\Gamma_{nm}+i k A_{nm}  + k^2 D_{nm}\right)\delta q_m(k) = \Lambda(k) \delta q_m(k),
\end{equation}
where summation over the repeated indices is assumed. We approach the problem perturbatively assuming that $k$ is small. The spectrum consists then of gapped and gapless modes. The value of the gap is determined by the smallest non-zero eigenvalue of $\Gamma$. The gapped modes decay in time and become irrelevant at sufficiently large time scales. On the other hand, there are also gapless modes that are related to the collision invariants of $\Gamma$. In the logic of the perturbation theory in $k$ the spectrum of the gapless modes has then the following structure
\begin{equation}
	\Lambda(k) = a  k + \omega k^2 + \mathcal{O}(k^3),
\end{equation}
and importantly the linear coefficients depend only on $A$ and therefore are universal, whereas the second order coefficients depend on both the collision integral and the GHD diffusion. The dependence on $\Gamma$ enters through the second-order perturbation theory from coupling of the zero-modes with the dispersive ones. On the other hand, the dependence on $D$, comes directly from the first order perturbation theory because of the $k^2$ factor standing in front of it. 

For the purpose of perturbation theory calculations it is convenient to introduce bra-ket notation such that $\Gamma_{nm} = \langle h_n| \Gamma | h_m \rangle$ and similarly for the other operators. More concretely, 
\begin{equation}
	\langle f|g \rangle = \int {\rm d}\lambda f(\lambda) g(\lambda), \qquad K| f\rangle = \int {\rm d}\lambda\, K(\cdot, \lambda) f(\lambda),
\end{equation}
with matrix elements understood in this notation as
\begin{equation}
    A_{nm}= \langle h_n|A |h_m \rangle = \langle h_n | \mathcal{A} \mathcal{C} |h_m \rangle, \qquad D_{nm} = \langle h_n | D | h_m \rangle= \frac{1}{2} \langle h_n | \mathfrak{D} \mathcal{C} | h_m \rangle, \qquad  \Gamma_{nm}= \langle h_n | \Gamma | h_m \rangle.
\end{equation}
Where $\mathcal{A}= (1-n \mathcal{T})^{-1} v (1-n \mathcal{T})$~\cite{lecture_notes_GHD,DrudeWeight_LL} and operators $\mathcal{C}, \mathfrak{D}$ are discussed in Sec.~\ref{secSM:sec1}.  Operator $\Gamma$ is discussed in Sec.~\ref{secSM:sec5} with a concrete example in \ref{secSM:sec9}. Moreover, for the basis $h_n$ introduced in Sec.~\ref{secSM:sec5}, the orthonormality condition is $\langle h_n | \mathcal{C} | h_m \rangle= \delta_{n,m}$.

With this notation, $|h_i\rangle$ with $i=0,1,2$ are eigenstates of $\Gamma$ with zero eigenvalues. We denote by $|g_\beta\rangle$ the eigenstates with non-zero eigenvalue $\gamma_\beta \neq 0$, $\Gamma |g_\beta\rangle = \gamma_\beta |g_\beta\rangle$. Whenever summation over $\beta$ appears, it means summation over eigenvectors with non-zero eigenvalue. On the other hand, index $\alpha$ always corresponds to thermal and sound modes.
\subsection{Perturbation theory at order $k$}
At this order we can neglect the GHD diffusion. The linear part of the dispersion relation of the hydrodynamic modes is then found by solving the eigenproblem in the degenerate zero subspace of the collision integral. The problem reduces to diagonalizing the following matrix
\begin{equation}\label{eqSM:Atrunc}
i k A_{\rm trunc}=
    i k\begin{pmatrix}
    0 & A_{1,0} & 0 \\
    A_{1,0} & 0 & A_{2,1} \\
    0 & A_{2,1} &0
    \end{pmatrix}.
\end{equation}
We find in total three modes with eigenvalues $\Lambda_{\pm}=\pm i v_s k$ (sound modes), and $\Lambda_{\rm th}=0$ (heat mode). The sound velocity given by the following formula, in agreement with thermodynamic sound velocity, see Sec.~\ref{secSM:sec4},
\begin{equation}
\label{eqSM:sound}
    v_s= \sqrt{A_{1,0}^2+A_{2,1}^2}= 1/\sqrt{\varrho \varkappa_S},
\end{equation}
and associated eigenvectors are (normalization factor is $\mathcal{N}= \sqrt{1+ (A_{2,1}/A_{1,0})^2}$)
\begin{equation}\label{eqSM:eigenvectors1st}
    |h_\pm\rangle =\frac{1}{\sqrt{2} \mathcal{N}} \bigg( \frac{A_{1,0}}{A_{2,1}} |h_0\rangle \pm \mathcal{N} |h_1\rangle + |h_2\rangle \bigg), \qquad
    |h_{\rm th}\rangle = \frac{1}{\mathcal{N}} \bigg(-|h_0\rangle + \frac{A_{1,0}}{A_{2,1}} |h_2\rangle \bigg).
\end{equation}
Therefore, in the first order of the perturbation theory, the degeneracy has been completely lifted. It is worth emphasizing here that the found eigenvalues (yielding the speed of sound) are not eigenvalues of the full matrix $A$. These are eigenvalues of matrix $A_{\rm trunc}$ truncated to subspace of charges ${q_0,q_1,q_2}$ conserved by the perturbation $\Gamma$.
Observe that eigenvectors \eqref{eqSM:eigenvectors1st} may be written entirely in terms of the specific heats, using the relations between hydrodynamic matrices and thermodynamics derived in Sec.~\ref{secSM:sec4},
\begin{equation}
\label{eqSM:modes_first}
    |h_{\pm}\rangle = \frac{1}{\sqrt{2}} \bigg( \sqrt{\frac{c_V}{c_P}} |h_0\rangle \pm |h_1 \rangle+ \sqrt{\frac{c_P-c_V}{c_P}}|h_2 \rangle \bigg),
    \qquad  |h_{\rm th}\rangle= - \sqrt{\frac{c_P-c_V}{c_P}}|h_0\rangle+\sqrt{\frac{c_V}{c_P}} |h_2 \rangle.
\end{equation}
This result can be compared with known formulas for the modes in the classical three-dimensional ideal gas. For the ideal gas $c_P/c_V=5/3$ and our formulas above reproduce results for sound and heat modes, when these values are substituted, see~\cite{Balescu1975}. Furthermore using~\eqref{eqSM:modes_first}, we also find useful relations for the sound modes in terms of the heat mode
\begin{equation} \label{SM:decomposition}
    | h_\pm \rangle = \sqrt{\frac{c_P-c_V}{2c_V}} |h_{\rm th} \rangle + \sqrt{\frac{c_P}{2 c_V}}  |h_0\rangle  \pm \frac{1}{\sqrt{2}}|h_1 \rangle,
\end{equation}
and heat mode in terms of ultra-local basis functions
\begin{equation}
\label{eqSM:heatultraloc}
    |h_{\rm th}\rangle= \frac{1}{\sqrt{\varrho T^2 c_P}} \left( |\mathscr{h}_2\rangle- \frac{\mathcal{B}_{2,1}}{\mathcal{B}_{1,0}}|\mathscr{h}_0\rangle \right).
\end{equation}
\subsubsection*{Gapless modes are equal to the modes of Navier-Stokes equations}
We complete considerations at this level of perturbation theory by noting that the found heat and sound modes coincide with the modes of Navier-Stokes equations \eqref{eqSM:NSheat&sound}.
For the purpose of this part, the bra-ket notation (which simplifies perturbation theory calculations) is not the most convenient. We therefore explicitly write sound and heat modes as 
\begin{equation}
\label{eqSM:heatsound}
    \delta q_\pm=\frac{1}{\sqrt{2}} \bigg( \sqrt{\frac{c_V}{c_P}} \delta q_0\pm \delta q_1+ \sqrt{\frac{c_P-c_V}{c_P}}\delta q_2 \bigg), \quad 
    \delta q_{\rm th}= -\sqrt{\frac{c_P-c_V}{c_P}}\delta q_0+\sqrt{\frac{c_V}{c_P}} \delta q_2.
\end{equation}
Already here we can see that these are exactly the modes of Navier-Stokes equation \eqref{eqSM:NSheat&sound}, after the following identification is made
\begin{equation}\label{eqSM:fieldrelations}
    \delta q_0 = \delta \tilde{\varrho}, \qquad \delta q_1= \delta \tilde{u}, \qquad \delta q_2 = \delta \tilde{T}.
\end{equation}
We can also observe the truncated matrix \eqref{eqSM:Atrunc} expressed with thermodynamic quantities \eqref{eqSM:thermoAmatrix} coincides with the relevant matrix \eqref{eqSM:NSmatrix} for linearized Navier-Stokes equations, when terms of order $k^2$ are neglected (these terms do not influence the eigenvectors in $k \to 0$ limit)
\begin{equation}
    ikA_{\rm trunc}=ik\begin{pmatrix}
    0 &  \frac{1}{\sqrt{\varrho \varkappa_T}}  & 0 \\
      \frac{1}{\sqrt{\varrho \varkappa_T}}  & 0 &  \sqrt{\frac{T}{\varrho^2 c_V}} \Partial{P}{T}{\varrho}  \\
    0 &  \sqrt{\frac{T}{\varrho^2 c_V}} \Partial{P}{T}{\varrho}  &0
    \end{pmatrix}.
\end{equation}
The relation \eqref{eqSM:fieldrelations} can also be justified from the very definition of the charges $q_0, q_1, q_2$.
To show this, we start with perturbations to ultra-local charges $\delta \mathscr{q}_0, \delta \mathscr{q}_1, \delta \mathscr{q}_2$, see \eqref{eqSM:charges012} and find
\begin{equation}
    \delta \mathscr{q}_0 = \sqrt{\mathcal{C}_{0,0}} \delta q_0, \qquad 
    \delta \mathscr{q}_1 = \sqrt{\mathcal{C}_{1,1}} \delta q_1, \qquad 
    \delta \mathscr{q}_2 = \sqrt{\frac{\mathcal{C}_{2,2} \mathcal{C}_{0,0}- \mathcal{C}_{2,0}^2}{\mathcal{C}_{0,0}}} \delta q_2 + \frac{\mathcal{C}_{2,0}}{\sqrt{C_{0,0}}} \delta q_0. 
\end{equation}
Perturbations of ultra-local charges can be easily connected to perturbations of the fields $\delta \varrho, \delta u, \delta e$ via relations \eqref{eqSM:udef} and \eqref{eqSM:edef}
\begin{equation}
    \delta \varrho = \delta \mathscr{q}_0 \qquad \delta u = \frac{1}{\varrho} \delta \mathscr{q}_1 \qquad \delta e = \frac{1}{\varrho} \delta \mathscr{q}_2 - \frac{\mathscr{q}_2}{\varrho^2} \delta \mathscr{q}_0.
\end{equation}
Lastly, to go to $\delta \varrho, \delta u, \delta T$ we need \eqref{eqSM:thermID}. Combining all three transformations together with thermodynamic formulas from Sec.\ref{secSM:sec4} (in particular, \eqref{eqSM:additionalthermo}), we find
\begin{equation}
	\delta \varrho = \sqrt{T \varrho \left(\frac{\partial P}{\partial \varrho} \right)_T^{-1}}\, \delta q_0=\varrho\sqrt{T \varkappa_T} \,\delta q_0 , \qquad \delta u = \sqrt{T/\varrho}\, \delta q_1, \qquad \delta T = T/\sqrt{\varrho c_V}\, \delta q_2.
\end{equation}
These relations are obviously equivalent to \eqref{eqSM:fieldrelations}. Summing up, we have shown that perturbations of charges $\delta q_0, \delta q_1, \delta_2$ are equal to perturbations of rescaled hydrodynamic fields.

\subsection{Perturbation theory at order $k^2$}
 We move now to calculations at the $k^2$ order. As already seen in \eqref{eqSM:SM_spectrum_NS}, at this level we should expect transport coefficients $\zeta, \kappa$ appearing in corrections to the heat and sound modes. The perturbation theory will provide us with explicit expressions for transport coefficients.
 
 The corrections of the order $k^2$ have two origins. There are second-order corrections coming from the coupling between the hydrodynamic modes and dispersive modes through the collision integral; we denote them $\omega_{\alpha}^{\rm coll}$. There are also corrections coming from the GHD diffusion denoted $\omega_{\alpha}^{\rm diff}$. Here $\alpha = \pm, {\rm th}$ specifies one of the hydrodynamic modes. Thus, the general structure is
 \begin{equation}
 \begin{aligned}
 	\Lambda_{\rm th}(k) &= \left(\omega_{\rm th}^{\rm diff} + \omega_{\rm th}^{\rm coll} \right) k^2 + \mathcal{O}(k^3), \\
	\Lambda_{\rm \pm}(k) &= \pm i v_s k + \left(\omega_{\rm \pm}^{\rm diff} + \omega_{\rm \pm}^{\rm coll} \right) k^2 + \mathcal{O}(k^3).
 \end{aligned}
 \end{equation}
The corrections due to the diffusion take a simple form. They are given by the first order perturbation theory in the diffusion operator $\omega_{\alpha}^{\rm diff} = \langle h_{\alpha} | D | h_{\alpha}\rangle$. Because the diffusion operator is zero between eigenstates of different parity and $|h_0\rangle$ is its zero eigenvector (both left and right) there are only $2$ non-zero matrix elements $D_{1,1}$ and $D_{2,2}$. In terms of them we find
\begin{equation}
\begin{aligned}
	\omega_{\rm th}^{\rm diff} &=  \frac{c_V}{c_P} D_{2,2}, \\
	\omega_{\pm}^{\rm diff} &= \frac{1}{2}D_{1,1} + \frac{1}{2} \frac{c_P - c_V}{c_P} D_{2,2}.
\end{aligned}
\end{equation}
Using the decomposition~\eqref{SM:decomposition} we also find
\begin{equation}
	\omega_{\pm}^{\rm diff} = \frac{c_P - c_V}{2 c_V}\omega_{\rm th}^{\rm diff} + \frac{1}{2} D_{1,1}.
\end{equation}

The second-order correction to the eigenvalues of the hydrodynamic modes come from the standard formula from the perturbation theory 
\begin{equation}
    \omega_\alpha^{\rm coll}=\sum_\beta  \frac{\langle h_\alpha| A| g_\beta \rangle \langle g_\beta| A |h_\alpha \rangle}{\gamma_\beta} , \qquad \alpha = \pm, {\rm th},
\end{equation}
where the sum does not include the kernel of operator $\Gamma$. 
Inspecting the formula we find $\omega_+^{\rm coll} = \omega_-^{\rm coll}$. This follows from the fact, that $\Gamma$ operator has eigenfunctions which are either symmetric or antisymmetric functions of rapidities.

Note that action of $A$ on $|h_0\rangle$ gives a mode proportional to $|h_1\rangle$  (this can be verified directly at the level of matrix elements \eqref{eqSM:matrixelements}) and therefore the middle term from the decomposition~\eqref{SM:decomposition} does not contribute to $\omega_{\pm}^{\rm coll}$ because $\langle g_\beta|h_1\rangle = 0$ from the definition of $| g_\beta \rangle$. In the expression for $\omega_{\pm}^{\rm coll}$ there appear also mixed terms of the form $\langle h_{\rm th}| A | g_{\beta}\rangle \langle g_\beta | A | h_1\rangle$. Such terms vanish because $A$ is non-zero only for states of different parity. This allows us to write
\begin{equation} \label{eqSM:omega_coll}
	\omega_{\pm}^{\rm coll} = \frac{c_P - c_V}{2 c_V} \omega_{{\rm th}}^{\rm coll} + \frac{1}{2} \omega_{1}^{\rm coll}, \qquad  \omega_{1}^{\rm coll} = \sum_\beta \frac{(\langle g_\beta | A | h_1 \rangle)^2 }{\gamma_\beta}.
\end{equation} 
\begin{figure}
    \centering
    \includegraphics[scale=0.5]{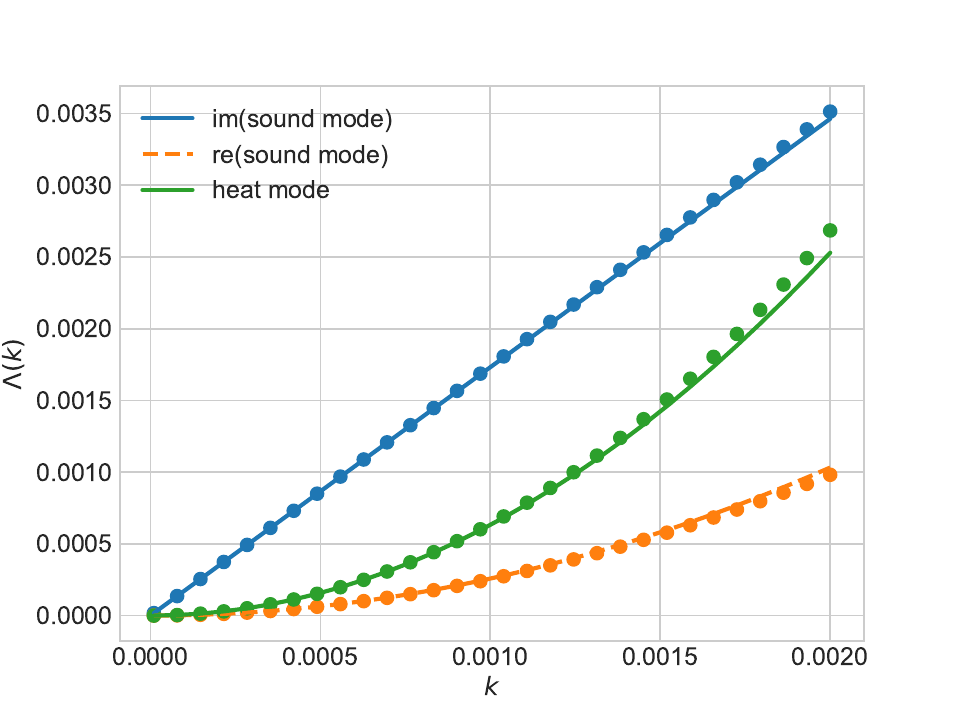}
    \includegraphics[scale=0.5]{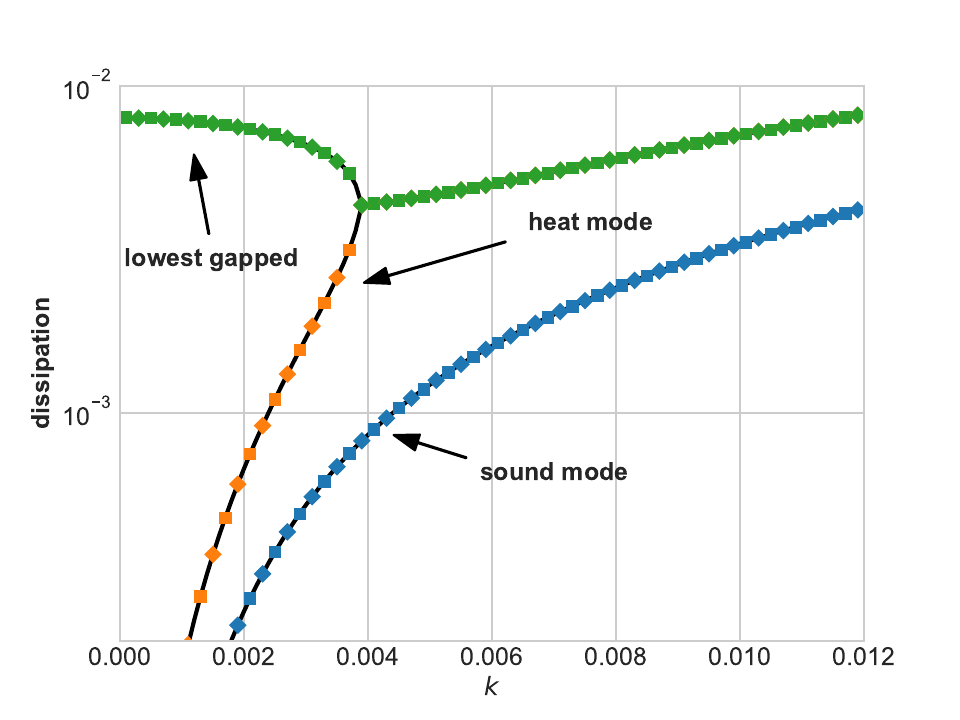}
    \caption{We present numerical solutions of~\eqref{eqSM:transporteigenproblem} for the coupled Lieb-Liniger model analyzed in the main text. The presented results are for $T=2$, $n=1$, $c=2$ and $\tau = 0.01$. {\em Left:} small-$k$ part of the spectrum of dispersion relations. For sufficiently small $k$, numerical results (points) agree with universal Navier-Stokes dispersion relations~\eqref{eqSM:SM_spectrum_NS} with transport coefficients~\eqref{eqSM:transportcoeffs} (lines). At larger momenta we see deviations caused by the higher order corrections in~$k$. We have multiplied the real parts of the dispersions by factor $5$ to make them of the same order as the imaginary parts. {\em Right:} a numerical solution requires truncating the infinite system of equations to a finite one. Truncating to first $12$ (squares), $18$ (diamonds) and $24$ (solid lines) charges yields results which differ on a scale below the resolution of the plot.}
    \label{figSM:transportc}
\end{figure}

\subsection{Transport coefficients}
We can finally read off the transport coefficients by comparing spectra~\eqref{eqSM:SM_spectrum_NS} of the linearized Navier-Stokes with the results of the perturbation theory. We find
\begin{equation}\label{eqSM:transportcoeffs}
\begin{aligned}
	\kappa^{\rm coll} &= \varrho c_P \omega_{\rm th}^{\rm coll}, \qquad \zeta^{\rm coll} =  \varrho \omega_1^{\rm coll}, \\
	\kappa^{\rm diff} &= \varrho c_V D_{2,2}, \qquad \zeta^{\rm diff} = \varrho D_{1,1}.
\end{aligned}
\end{equation}
with $\kappa = \kappa^{\rm coll} + \kappa^{\rm diff}$ and $\zeta = \zeta^{\rm coll} + \zeta^{\rm diff}$. 

\begin{figure}
    \centering
    \includegraphics[scale=0.5]{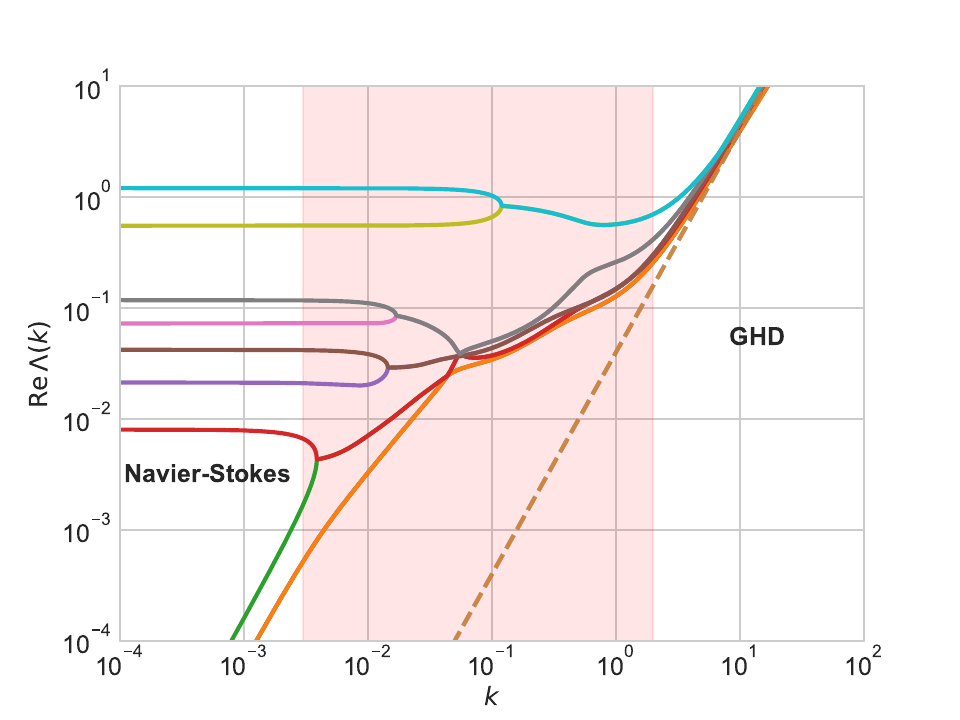}
    \caption{The real part of the dispersion relations for the lowest lying modes from the exact numerical solution of~\eqref{eqSM:transporteigenproblem}. At small momenta, the spectrum is characterized by the $3$ gapless modes (the orange curve is doubly degenerate) of the Navier-Stokes regime whereas at (relatively) large momenta by the modes of the GHD. The dashed lines are the gapless modes of the GHD without the integrability breaking terms. The joining and splitting of the modes defines an intermediate regime (shaded) where the dynamics cannot be reduced to either Navier-Stokes or unperturbed GHD.}
    \label{figSM:transportfullk}
\end{figure}

We confirm these formulas on Fig.\ref{figSM:transportc}, which presents small-$k$ part of the spectrum found in numerical diagonalization of \eqref{eqSM:transporteigenproblem}. Our results agree with numerics in the samll-$k$ regime, as expected. The dispersion relations for larger number of modes and for higher values of $k$ are presented in Fig.\ref{figSM:transportfullk}. For small $k$ the spectrum consists of Navier-Stokes modes and gapped modes. For high enough $k$, the spectrum is those of unperturbed GHD. This corresponds to operator $\Gamma$ being negligible in the eigenproblem \eqref{eqSM:transporteigenproblem}, as compared to $A$ and $D$. In between, there is a region which cannot be described neither by Navier-Stokes hydrodynamics, nor the pure diffusive GHD. Therefore, by numerically solving \eqref{eqSM:transporteigenproblem} we are able to quantatively capture the crossover from Navier-Stokes hydrodynamics to GHD.

Equations~\eqref{eqSM:transportcoeffs} together with~\eqref{eqSM:omega_coll} provide also an efficient way to evaluate the transport coefficients numerically. The procedure is the following. For a given thermal equilibrum state, we construct an orthonormal basis $\{|h_i\rangle\}$ using the hydrodynamic scalar product $(f|g)$. This allows us to compute the matrix elements $A_{nm}$, $\Gamma_{nm}$ and $D_{nm}$. The matrix $\Gamma_{nm}$ can be now diagonalized numerically. We obtain the spectrum $\gamma_{\beta}$ but also expressions for $|g_\beta\rangle$ in terms of linear combinations of $|h_i\rangle$'s. This allows us to evaluate the matrix elements $\langle g_\beta| A | h_{\alpha}\rangle$ and $\omega_{\alpha}^{\rm coll}$. The remaining ingredient is the heat capacity $c_P$ which can be computed from the hydrodynamic matrices $\mathcal{B}$ and $\mathcal{C}$ according to~\eqref{eqSM:c_P}. The numerical solution requires truncating the infinite basis to a finite one. In Fig.~\ref{figSM:transportc} we show that it is sufficient to take around $20$ basis functions.

Let us note that contributions $\kappa^{\rm diff}$ and $\zeta^{\rm diff}$ can be also written as matrix elements in the ultra-local basis, as in the main text. Using relations \eqref{eqSM:charges012}, properties of the diffusion kernel and adopting notation of matrix elements of diffusion operator \eqref{eqSM:matrixelements} as in the main text we find that indeed
\begin{equation} \label{transport_diff_ultra}
    \kappa^{\rm diff}= \frac{1}{2T^2}\mathscr{h}_2 \mathfrak{D} \mathcal{C} \mathscr{h}_2, \qquad \zeta^{\rm diff}= \frac{1}{2T}\mathscr{h}_1 \mathfrak{D} \mathcal{C} \mathscr{h}_1.
\end{equation}
Importantly $\kappa^{\rm diff}, \zeta^{\rm diff} \ge 0$, due to the positivity of operator $\mathfrak{D}\mathcal{C}$~\cite{hydro_diff_prl}. Moreover, it is easy to observe that $\omega_1^{\rm coll}, \omega^{\rm coll}_{\rm th} \ge 0$ and hence $\kappa^{\rm coll}, \zeta^{\rm coll} \ge 0$. We thus conclude that transport coefficients found by us are non-negative, which assures that entropy production in the Navier-Stokes dynamics is positive~\cite{Balescu1975}.

In the next section we will confirm that $\kappa^{\rm coll}$ and $\zeta^{\rm coll}$ found in the second-order perturbation theory \eqref{eqSM:transportcoeffs} can be written as generalized Chapman-Enskog expressions given in the main text.

\section{Integral equations for $\kappa^{\rm coll}$ and $\zeta^{\rm coll}$ transport coefficients}\label{secSM:sec8}
Expressions for the collisional transport coefficients involve sums over all non-zero eigenstates of the $\Gamma$ operator. In the following, we will show how they can be rewritten as integral equations, which are identical to the integral equations found in the Chapman-Enskog procedure.
We start with the thermal conductivity related to $\omega_{\rm th}^{\rm coll}$. As mentioned earlier for any $\beta$ we have $\langle h_0|A|g_\beta\rangle=0$. Using explicit form \eqref{eqSM:heatultraloc} of $|h_{\rm th} \rangle$ in ultra-local basis we find 
\begin{equation}
    \langle h_{\rm th}|A|g_\beta \rangle = \frac{1}{\sqrt{\varrho T^2 c_P}} \langle \mathscr{h}_2|A|g_\beta \rangle,
\end{equation}
and thus we have
\begin{equation}
    \omega_{\rm th}^{\rm coll}=\frac{1}{\sqrt{\varrho T^2 c_P}} \sum_\beta \frac{\langle \mathscr{h}_2|A|g_\beta \rangle \langle g_\beta |A|h_{\rm th} \rangle}{\gamma_\beta}.
\end{equation}

For the heat mode the sum can be easily extended to the whole space by noting that $A|h_{\rm th}\rangle$ is orthogonal to the kernel of $\Gamma$. For $|h_0 \rangle, |h_2 \rangle$ it is clear from symmetry properties of $A$. For $|h_1 \rangle$ we can show $\langle h_1 |A| h_{\rm th} \rangle=0$, which be seen from explicit computation using \eqref{eqSM:heatultraloc}. This allows us to write
\begin{equation}
    \omega_{\rm th}^{\rm coll} = \lim_{\epsilon \to 0}\frac{1}{\sqrt{\varrho T^2 c_P}} \langle \mathscr{h}_2| A \Gamma^{-1}_\epsilon A |h_{\rm th} \rangle,
\end{equation}
where we used the completeness of the basis which implies that 
\begin{equation}
    \lim_{\varepsilon \to 0} \sum_m | g_m \rangle \frac{1}{\gamma_m-\varepsilon} \langle g_m | = \lim_{\varepsilon \to 0} \frac{1}{\Gamma-\varepsilon \mathcal{C}} \equiv \lim_{\epsilon \to 0} \Gamma^{-1}_\epsilon,
\end{equation}
where $m$ extends now over all the eigenstates. 
Now, we can formally introduce vector $| \phi_{\kappa}\rangle$ such that,
\begin{equation}
    \Gamma_\epsilon^{-1}A |h_{\rm th}\rangle =\frac{T}{\sqrt{\varrho c_P}} | \phi_\kappa \rangle, \qquad \omega_{\rm th}^{\rm coll} = \frac{1}{\varrho c_P} \langle \mathscr{h}_2 |A |\phi_{\kappa}\rangle.
\end{equation}
To find $|\phi_{\kappa} \rangle$ we can go to quasiparticle representation. We look for a function $\eta_\kappa(\lambda)$ which fulfills the following integral equation
\begin{equation}\label{eqSM:Gammaeps}
    \Gamma_\epsilon \phi_{\kappa}= (1-n\mathcal{T})^{-1}\rho_{\rm p} f \eta_{\kappa},
\end{equation}
with function $\eta_{\kappa}(\lambda)$ related to heat mode and given by
\begin{equation}\label{eqSM:etakappa}
    \eta_\kappa(\lambda) = \frac{\sqrt{\varrho c_P}}{T} v(\lambda) h_{\rm th}^{\rm dr}(\lambda)= \frac{\sqrt{\varrho c_V}}{T} \left( h_2 - \sqrt{\frac{c_P-c_V}{c_V}} h_0 \right)^{\rm dr}(\lambda).
\end{equation}
For $\epsilon \neq 0$, the operator $\Gamma_\epsilon$ is invertible and the solution to \eqref{eqSM:Gammaeps} is unique. However, we would like to explicitly take the limit $\epsilon \to 0$, which will invalidate uniqueness of the solution. This is clear from the existence of collision invariants of $\Gamma$ - for any particular solution $\bar{\phi}_{\kappa}$ to the equation \eqref{eqSM:Gammaeps} with $\epsilon=0$ the function 
\begin{equation}
    \bar{\phi}_{\kappa}+ \sum_{i=0}^2 a_i \mathscr{h}_i
\end{equation}
is an another solution (coefficients $a_i$ are arbitrary). Therefore, before we set $\epsilon=0$, let us  multiply \eqref{eqSM:Gammaeps} with collision invariants $\mathscr{h}_i(\lambda)$ where $i=0,1,2$ and integrate over $\dd \lambda$. This leads to the conditions
\begin{equation}
    \langle \mathscr{h}_i|\mathcal{C}| \phi_{\kappa} \rangle=(\mathscr{h}_i|\phi_{\kappa})=0,
\end{equation}
which we keep after taking $\epsilon \to 0$. The conditions above make the solution unique by fixing the constants $a_i$ in the general solution.

In the main text we shorten the notation introducing $\chi_\kappa=(1-n\mathcal{T})^{-1}\rho_{\rm p} f \eta_{\kappa}$, which explicitly reads
\begin{equation}
    \chi_\kappa= \frac{\sqrt{\varrho c_V}}{T} \mathcal{B}(h_2 - \sqrt{c_P/c_V-1} \, h_0). 
\end{equation}
The integral equation can be then written as $\Gamma \phi_\kappa=\chi_\kappa$, as in the main text. Finally, thermal conductivity $\kappa^{\rm coll}$ is given by 
\begin{equation}
    \kappa^{\rm coll} = \int \dd \lambda \rho_{\rm p}(\lambda) f(\lambda) v(\lambda) \mathscr{h}_2^{\rm dr}(\lambda) \phi_\kappa^{\rm dr}(\lambda) = \mathscr{h}_2 \mathcal{B} \phi_\kappa.
\end{equation}
We focus now on $\zeta^{\rm coll}$. We proceed similarly, with the exception that now $\langle g_\beta|A|h_1\rangle \neq 0$, therefore we have to properly regularize matrix elements before we introduce representation of inverse $\Gamma$. We start with expression
\begin{equation}
    \langle g_\beta |A| h_1 \rangle,
\end{equation}
which enters the sum for $\zeta^{\rm coll}$. We can add to it two zero terms, namely:
\begin{equation}
    \langle g_\beta |A| h_1 \rangle + a \langle g_\beta |h_0 \rangle + b \langle g_\beta |h_2 \rangle.
\end{equation}
We want $a,b$ to be such that the vector $|w_1 \rangle =A |h_1 \rangle + a |h_{0} \rangle + b |h_2 \rangle$ is orthogonal to $|h_0\rangle, |h_1\rangle, |h_2\rangle$. We find $a = -A_{1,0}, b= -A_{2,1}.$ Thus, we write (we have used here that $|h_1 \rangle = 1/\sqrt{\varrho T} |\mathscr{h}_1 \rangle$, see \eqref{eqSM:charges012} and Sec.\ref{secSM:sec4})
\begin{equation}
	 \omega_{1}^{\rm coll} = \frac{1}{\sqrt{\varrho T}}\lim_{\epsilon\rightarrow 0} \langle \mathscr{h}_1 | A \Gamma_\epsilon^{-1} |w_1 \rangle.
\end{equation}
As previously, we introduce a vector $|\phi_\zeta \rangle$ such that $\Gamma_\epsilon^{-1}|w_1\rangle= \sqrt{T/\varrho} |\phi_\zeta \rangle $. In quasiparticle representation finding such vector corresponds to solving the following integral equation
\begin{equation}
    \Gamma_\epsilon \phi_{\zeta}= (1-n\mathcal{T})^{-1}\rho_{\rm p} f \eta_{\zeta},
\end{equation}
with function (we have used here \eqref{eqSM:thermoAmatrix})
\begin{equation}\label{eqSM:etazeta}
    \eta_\zeta(\lambda) = \sqrt{\frac{\varrho}{T}} \left[ v(\lambda)h_1^{\rm dr}(\lambda) - \frac{1}{\sqrt{\varrho \varkappa_T}} \left( h_0 + \sqrt{\frac{c_P-c_V}{c_V}}h_2 \right)^{\rm dr}(\lambda) \right]. 
\end{equation}
Similarly as for thermal conductivity we set $\epsilon \to 0$ by imposing additional conditions $(\mathscr{h}_i|\phi_\zeta)=0, \, \, i=0,1,2$. We shorten the notation and introduce $\chi_\zeta= (1-n\mathcal{T})^{-1}\rho_{\rm p} f \eta_{\zeta}$, which explicitly reads
\begin{equation}
    \chi_\zeta= \sqrt{\frac{\varrho}{T}} \mathcal{B} h_1 - \frac{1}{\sqrt{T \varkappa_T}} \, \mathcal{C} (h_0+\sqrt{c_P/c_V-1}\,h_2).
\end{equation}
The integral equation is then $\Gamma \phi_\zeta = \chi_\zeta$ and the bulk viscosity    $\zeta^{\rm coll}$ finally is
\begin{equation}
    \zeta^{\rm coll} = \int \dd \lambda \rho_{\rm p}(\lambda) f(\lambda) v(\lambda) \mathscr{h}_1^{\rm dr}(\lambda) \phi_\zeta^{\rm dr}(\lambda) = \mathscr{h}_1 \mathcal{B} \phi_\zeta.
\end{equation}
In this way, we have shown that results from perturbation theory can be reformulated as generalized Chapman-Enskog integral equations, given in the main text. One of the advantages of integral equation approach is simple analysis of the non-interacting limit of the underlying integrable model. This is done in the next section.

\subsection*{Non-interacting limit of transport coefficients}
In this section we look more closely on the limit of non-interacting integrable model with trivial dressings. Transport coefficients \eqref{eqSM:transportcoeffs} have contributions from collision integral and from diffusion operator. In the non-interacting limit, diffusion operator vanishes together with the associated contributions to the transport coefficients. We thus turn to contributions from collision integral, which in principle may be non-zero even in the limit of free integrable model. Obviously, we still consider non-zero integrability breaking term $\mathcal{I}$.

In the case of trivial dressings, the following symmetries appear on the level of hydrodynamic matrices
\begin{equation}
\label{eqSM:free_sym}
    \mathcal{B}_{1,0}=2 C_{2,0}, \qquad \mathcal{B}_{2,1}=2\mathcal{C}_{2,2}.
\end{equation}
Let us look how these affect expressions for $\eta_{\kappa,\zeta}(\lambda)$ given by \eqref{eqSM:etakappa}, \eqref{eqSM:etazeta}. It is convenient to represent these functions in the ultra-local basis first, we find
\begin{align}
    \eta_\kappa(\lambda) =& \frac{1}{T^2}v(\lambda) \left( \mathscr{h}_2^{\rm dr}(\lambda) - \frac{\mathcal{B}_{2,1}}{\mathcal{B}_{1,0}}\mathscr{h}_0^{\rm dr}(\lambda) \right),\\
    \eta_\zeta (\lambda) =& \frac{1}{T} \left(v(\lambda) \mathscr{h}^{\rm dr}_1(\lambda) +\frac{\mathcal{B}_{2,1}\mathcal{C}_{2,0}-\mathcal{B}_{1,0}\mathcal{C}_{2,2}}{\mathcal{C}_{2,2}\mathcal{C}_{0,0}-\mathcal{C}_{2,0}^2}\mathscr{h}_0^{\rm dr}(\lambda) -  \frac{\mathcal{B}_{2,1}\mathcal{C}_{0,0}-\mathcal{B}_{1,0}\mathcal{C}_{2,0}}{\mathcal{C}_{2,2}\mathcal{C}_{0,0}-\mathcal{C}_{2,0}^2}\mathscr{h}_2^{\rm dr}(\lambda)\right),
\end{align}
where we have expressed thermodynamic quantities with hydrodynamic matrices, see Sec.~\ref{secSM:sec4}. Moreover, we have represented GS-orthonormalized vectors \eqref{eqSM:charges012} with ultra-local counterparts. Let us consider a limit, where dressing can be neglected and hence $v(\lambda)=\mathscr{h}_1(\lambda)$. Taking into account also the symmetries \eqref{eqSM:free_sym}, we observe
\begin{align}
    \eta_\kappa(\lambda) \to & \, \frac{1}{T^2} \mathscr{h}_1(\lambda) \left( \mathscr{h}_2(\lambda) - \frac{\mathcal{B}_{2,1}}{\mathcal{B}_{1,0}}\mathscr{h}_0(\lambda) \right),\\
    \eta_\zeta (\lambda) \to& \, 0.
\end{align}
The rate on which $\eta_\zeta (\lambda)$ approaches zero with vanishing coupling depends on the model. This might be important in cases where $\Gamma$ vanishes in the free limit as well. When $\eta_\zeta(\lambda)$ approaches zero faster than $\Gamma$ with the interaction coupling, the bulk viscosity tends to 0. This is the case for collision integral studied by us. On the other hand, as $\eta_\kappa(\lambda)$ remains finite in the non-interacting limit, the $\Gamma$ operator, which vanishes in the free limit leads then to infinite thermal conductivity. This is a known result for non-interacting gas~\cite{Balescu1975} and we observe it also for collision integral considered by us (see Fig.~2 in the main text).

\section{Collision integral for the coupled Lieb-Liniger models}\label{secSM:sec9}

In this supplementary material we derive the linearized collision integral for the two coupled Lieb-Liniger models. The tubes are coupled with intertube interaction potential $V_{\rm inter}(x-y)=V_0 \, V(x-y).$ The resulting collision integral has all the necessary properties which, we have assumed about the $\Gamma$ operator and will allow us to illustrate our general formalism in an experimentally relevant situation.

Generally speaking there are two contributions to the collision integral $\mathcal{I}$. The direct contributions comes from the very scattering events caused by the perturbing potential. But there is also an indirect contribution, that arises because a modification to the distribution of rapidities in one place (in the rapidity space) reshuffles rapidities elsewhere due to the interaction present in the Lieb-Liniger model. The two contributions together appear as a dressing of the bare scattering integral $\mathcal{I}_0$
\begin{equation}
	\mathcal{I}[\rho_{\rm{p},1}, \rho_{\rm{p},2}](\lambda) = \tau_0^{-1}\int \dd \mu \mathbb{F}(\lambda,\mu)  \mathcal{I}_0[\rho_{\rm{p},1}, \rho_{\rm{p},2}](\mu),
\end{equation}
where $\mathbb{F}$ is backflow operator introduced in \eqref{eqSM:Foperator} and timescale $\tau_0^{-1}=16 E_F/(\hbar \pi^2) \times \gamma_{\rm inter}^2 = \frac{2 V_0^2m}{\hbar^3}$ where $E_F$ is Fermi energy of a single tube and $\gamma_{\rm inter}=V_0 m /(\varrho \hbar^2)$ where $\varrho$ is the density of the gas~\cite{PGK}.

As discussed in~\cite{PGK,Lebek2024} in the homogeneous case when two tubes are in the same state the leading contribution to the collision integral comes from processes of type $(1,2)$ and $(2,1)$. These are processes in which $1$ particle-hole pair is created in the first tube and $2$ particle-hole pair is created in the second tube or {\em vice verse}, $\mathcal{I}[\rho_{\rm{p}}] = \mathcal{I}^{(1,2)}[\rho_{\rm{p}}] + \mathcal{I}^{(2,1)}[\rho_{\rm{p}}] + (\dots)$, where $(\dots)$ stands for processes of higher order in momentum transferred between the tubes. 

The bare collision integral for these processes is~\cite{PGK,Lebek2024}
\begin{equation}
\label{eqSM:mncontributionv2}
\begin{aligned}
    \mathcal{I}_0[\rho_{\rm p}](\lambda)=&\frac{(2 \pi)^2}{(2!)^2} \int \dd p_{1,1} \dd h_{1,1} \dd p_{1,2} \dd h_{1,2} \dd p_{2} \dd h_{2}  \, \left( \delta(\lambda-p_{1,1})+\delta(\lambda-p_{1,2}) + \delta(\lambda-p_{2}) \right) \,
    \tilde{V}^2\big(k_{2 \rm ph}(\mathbf{p}_1,\mathbf{h}_1)\big) 
     \times \\
    & |F_{2 \rm ph}(\mathbf{p}_1,\mathbf{h}_1)|^2 |F_{1 \rm ph}(\mathbf{p}_2,\mathbf{h}_2)|^2 \, \delta\big(k_{2 \rm ph}(\mathbf{p}_1,\mathbf{h}_1)+k_{1 \rm ph}(\mathbf{p}_2,\mathbf{h}_2)\big) \, \delta\big(\omega_{2 \rm ph}(\mathbf{p}_1,\mathbf{h}_1)+\omega_{1 \rm ph}(\mathbf{p}_2,\mathbf{h}_2)\big) \times \\
    &\Big(J_{2 \rm ph}(\mathbf{p}_1,\mathbf{h}_1) \, J_{1 \rm ph}(\mathbf{p}_2,\mathbf{h}_2)-(\mathbf{h} \leftrightarrow \mathbf{p}) \Big),
\end{aligned}
\end{equation}
where $\tilde{V}(k)$ is Fourier transform of the interaction potential with $\tilde{V}(k)=\tilde{V}(-k)$, $F_{n \rm ph}(\mathbf{p},\mathbf{h})$ are density operator form factors and
\begin{equation}
    k_{n \rm ph}(\mathbf{p},\mathbf{h})=\sum_{i=1}^n \left( k(p_i)-k(h_i)\right), \qquad \omega_{n \rm ph}(\mathbf{p},\mathbf{h})= \sum_{i=1}^n \left( \omega(p_i)-\omega(h_i) \right).
\end{equation}
where $k(\lambda)=\mathscr{h}_1^{\rm Dr}(\lambda)$ and $\omega (\lambda)= \mathscr{h}_2^{\rm Dr}(\lambda)$. Finally, $J$ factors are
\begin{equation}
    J_{n \rm ph}(\mathbf{p},\mathbf{h})=\prod_{i=1}^n\rho_{\rm p}(h_i)\rho_{\rm h}(p_i).
\end{equation}

We linearize now the collision integral around a thermal state. Any thermal state is a stationary state fulfilling the condition
\begin{equation}
    J_{2 \rm ph}(\mathbf{p}_1,\mathbf{h}_1) \, J_{1 \rm ph}(\mathbf{p}_2,\mathbf{h}_2)-J_{2 \rm ph}(\mathbf{h}_1,\mathbf{p}_1) \, J_{1 \rm ph}(\mathbf{h}_2,\mathbf{p}_2)=0,
\end{equation}
on a $4$-dimensional manifold given by $\omega_{2 \rm ph}(\mathbf{p}_1,\mathbf{h}_1)+\omega_{1 \rm ph}(\mathbf{p}_2,\mathbf{h}_2) = 0$ and $k_{2 \rm ph}(\mathbf{p}_1,\mathbf{h}_1)+k_{1 \rm ph}(\mathbf{p}_2,\mathbf{h}_2) = 0$. 
To expand the collision integral around the thermal state it is useful to switch from the distribution functions to the pseudoenergy $\epsilon(\lambda)$ for which we have
\begin{equation}
	\rho_{\rm p}(\lambda) = \frac{\rho_{\rm tot}(\lambda)}{1 + e^{\epsilon(\lambda)}}, \qquad \rho_{\rm h}(\lambda) = \frac{\rho_{\rm tot}(\lambda)}{1 + e^{-\epsilon(\lambda)}}.
\end{equation}
The ratio of the $J$-factors has then a simple form
\begin{equation}
        \frac{J_{2 \rm ph}(\mathbf{p}_1,\mathbf{h}_1) \, J_{1 \rm ph}(\mathbf{p}_2,\mathbf{h}_2)}{J_{2 \rm ph}(\mathbf{h}_1,\mathbf{p}_1) \, J_{1 \rm ph}(\mathbf{h}_2,\mathbf{p}_2)} = \exp \left(\epsilon(p_{1,1})-\epsilon(h_{1,1})+\epsilon(p_{1,2})-\epsilon(h_{1,2})+\epsilon(p_{2,1})-\epsilon(h_{2,1}) \right),
\end{equation}
and the argument of the exponential vanishes for thermal states for which $\epsilon(\lambda) = \beta_2 \omega(\lambda) + \beta_0$. 

To linearize the collision integral we consider a deviation from the thermal state $\epsilon \to \epsilon+\delta \epsilon $ and look at $\mathcal{I}[\epsilon +\delta \epsilon]$. To understand the structure of $\mathcal{I}[\epsilon +\delta \epsilon]$ we analyze first $\epsilon +\delta \epsilon$. From \eqref{eqSM:epsilonD} we conclude that 
\begin{equation}
    \epsilon'(\lambda)=(\epsilon_0')^{\rm dr}(\lambda).
\end{equation}
Using now the relation~\cite{PhysRevLett.124.140603} between large and small dressing operations $(f^{\rm Dr})'(\lambda)=(f')^{\rm dr}(\lambda)$ and integrating we find
\begin{equation}
    \epsilon(\lambda)= (\epsilon_0)^{\rm Dr}(\lambda) + const
\end{equation}
The same reasoning can be applied to the perturbed state, we get
\begin{equation}
    \epsilon(\lambda)+\delta \epsilon(\lambda) + \mathcal{O}((\delta \epsilon_0)^2) = (\epsilon_0+\delta \epsilon_0)^{ {\rm Dr}+\delta {\rm Dr}}(\lambda) +const,
\end{equation}
where by $ {\rm Dr}+\delta {\rm Dr}$ we denoted Dressing implemented by the state with occupation function $n+\delta n$. The expression above may be expanded according to 
\begin{equation}
    \epsilon(\lambda)+\delta \epsilon(\lambda)= \epsilon_0^{Dr +\delta Dr}(\lambda) + (\delta \epsilon_0)^{Dr +\delta Dr} +const  = \epsilon_0^{Dr+\delta Dr}(\lambda) +(\delta \epsilon_0)^{Dr}(\lambda) +const+ O((\delta \epsilon_0)^2).
\end{equation}
With this, we evaluate $\mathcal{I}[\epsilon+\delta \epsilon]$. The result is 
\begin{equation}
\label{eqSM:gam0}
    \mathcal{I}_0[\rho] = \Gamma_0[\rho_{\rm th}] (\delta \epsilon_0)^{\rm Dr} + \mathcal{O}((\delta\epsilon)^2),
\end{equation}
where
\begin{equation}
\label{eqSM:mncontributionv2}
\begin{aligned}
    \Gamma_0(\lambda,\mu)=&\frac{(2 \pi)^2}{8} \int \dd p_{1,1} \dd h_{1,1} \dd p_{1,2} \dd h_{1,2} \dd p_{2} \dd h_{2} \,  G(\bfp_1, \bfh_1, \bfp_2, \bfh_2)  \times \\
    &\Big(
    \delta (\lambda-p_{1,1})-\delta (\lambda-h_{1,1})+\delta (\lambda-p_{1,2})-\delta (\lambda-h_{1,2})+\delta (\lambda-p_{2})- \delta (\lambda-h_{2}) \Big) \times \\
    &\Big(
    \delta (\mu-p_{1,1})-\delta (\mu-h_{1,1})+\delta (\mu-p_{1,2})-\delta (\mu-h_{1,2})+\delta (\mu-p_{2})- \delta (\mu-h_{2})
 \Big),
\end{aligned}
\end{equation}
with
\begin{align}
	G(\bfp_1, \bfh_1, \bfp_2, \bfh_2) =& 
    \tilde{V}^2\big(k_{2 \rm ph}(\mathbf{p}_1,\mathbf{h}_1)\big) |F_{2 \rm ph}(\mathbf{p}_1,\mathbf{h}_1)|^2|F_{1 \rm ph}(\mathbf{p}_2,\mathbf{h}_2)|^2  J_{2 \rm ph}(\mathbf{p}_1,\mathbf{h}_1) \, J_{1 \rm ph}(\mathbf{p}_2,\mathbf{h}_2) \times \nonumber \\
    &\delta\big(k_{2 \rm ph}(\mathbf{p}_1,\mathbf{h}_1)+k_{1 \rm ph}(\mathbf{p}_2,\mathbf{h}_2)\big) \, \delta\big(\omega_{2 \rm ph}(\mathbf{p}_1,\mathbf{h}_1)+\omega_{1 \rm ph}(\mathbf{p}_2,\mathbf{h}_2)\big).
\end{align}
Note that $G(\bfp_1, \bfh_1, \bfp_2, \bfh_2) = G(\bfh_1, \bfp_1, \bfh_2, \bfp_2)$, namely function $G$ is $\rm ph$-symmetric. This implies that operator $\Gamma_0$ is symmetric, $\Gamma_0(\lambda,\mu)=\Gamma_0(\mu,\lambda)$ leading to  $\Gamma_{mn}=\Gamma_{nm}$. It is also easy to observe that $\Gamma_{nn} \geq 0$, hence $\Gamma$ is positive semi-definite.

What is more, $\Gamma_0$ has an additional symmetry. By performing change of variables in the integrals (changing sign in each one) we get $\Gamma_0(\lambda,\mu)=\Gamma_0(-\lambda,-\mu)$. This property leads to certain selection rules once we consider the action of $\Gamma$ in the basis of ultra-local conserved charges. 

The linearized collision integral has also three left eigenvectors with zero eigenvalues~\cite{Lebek2024}. They reflect the fact that the total number of particles, the total momentum and the total energy are conserved by the perturbation,
\begin{equation}
	\int {\rm d} \lambda \mathscr{h}_0^{\rm Dr}(\lambda) \Gamma_0(\lambda,\mu) = \int {\rm d} \lambda \mathscr{h}_1^{\rm Dr}(\lambda) \Gamma_0(\lambda,\mu) = \int {\rm d} \lambda \mathscr{h}_2^{\rm Dr}(\lambda) \Gamma_0(\lambda,\mu) = 0,
\end{equation}
where to derive the last two properties we used conservation laws under the integral.

We consider now matrix elements of $\Gamma$ in the basis of the ultra-local conserved charges. The matrix elements are defined according to eq.~\eqref{eqSM:matrixelements},
\begin{equation}
    \Gamma_{nm}= \tau_0^{-1}\int \dd \lambda \dd \mu \,  h_n^{\rm Dr}(\lambda) \Gamma_0(\lambda, \mu) h_m^{\rm Dr}(\mu).
\end{equation}

From the construction of our basis (see Sec.\ref{secSM:sec5}), functions $h_n$ are either symmetric or antisymmetric (and  dressings do not change this property). This implies that $\Gamma_{nm} = 0$ whenever functions $h_n$ and $h_m$ have different parity. According to the above discussion on the zero eigenvectors we also have $\Gamma_{nm} = 0$ for $n = 0, 1, 2$.  Moreover, it is straightforward to observe that $\Gamma_{nm} = 0$ for $m=0,1,2$ as well.

\subsection*{Small momentum limit}

The matrix elements $\Gamma_{nm}$ read explicitly
\begin{equation}
\label{eqSM:mncontributionv2}
\begin{aligned}
    \Gamma_{nm}=& \tau_0^{-1} \times\frac{(2 \pi)^2}{8} \int \dd p_{1,1} \dd h_{1,1} \dd p_{1,2} \dd h_{1,2} \dd p_{2} \dd h_{2}\,  G(\bfp_1, \bfh_1, \bfp_2, \bfh_2) \times \\
    &\Big( h_n^{\rm Dr}(p_{1,1})-h_n^{\rm Dr}(h_{1,1})+h_n^{\rm Dr}(p_{1,2})-h_n^{\rm Dr}(h_{1,2})+h_n^{\rm Dr}(p_{2})- h_n^{\rm Dr}(h_{2}) \Big) \times \\
    &\Big(
    h_m^{\rm Dr}(p_{1,1})-h_m^{\rm Dr}(h_{1,1})+h_m^{\rm Dr}(p_{1,2})-h_m^{\rm Dr}(h_{1,2})+h_m^{\rm Dr}(p_{2})- h_m^{\rm Dr}(h_{2})
 \Big).
\end{aligned}
\end{equation}
We will find now a much simpler expression for them in the small momentum limit.
To this end we express the integrand in terms of center of mass rapidities and the momentum transferred between the tubes. 

First, we change the integration variables to the center-of-mass rapidities. These are defined through $\lambda = (p+h)/2$ and $\alpha = p-h$ for each particle-hole excitation. The Jacobian of the transformation from $(p,h)$ to $(\lambda, \alpha)$ is $1$ hence this amounts to simply replacing the integration measure by $\dd \lambda_{1,1} \dd \lambda_{1,2} \dd \lambda_{2} \dd \alpha_{1,1} \dd \alpha_{1,2} \dd \alpha_2$. We then assume that the main contribution to the integral comes from the small momentum excitations. Therefore, the integrand can be expanded in small $\alpha$'s. Specifically, the energy-momentum constraints become
\begin{subequations} \label{momentum_energy_small}
    \begin{equation}
        k'(\lambda_{1,1})\alpha_{1,1}+k'(\lambda_{1,2})\alpha_{1,2}+k'(\lambda_{2})\alpha_{2}=0,
    \end{equation}
    \begin{equation}
        \omega'(\lambda_{1,1})\alpha_{1,1}+\omega'(\lambda_{1,2})\alpha_{1,2}+\omega'(\lambda_{2})\alpha_{2}=0.
    \end{equation}
\end{subequations}
Solving them for $\alpha_{1,1}$ and $\alpha_{1,2}$ gives
\begin{align}
    \bar{\alpha}_{1,1} = - \alpha_{2} \frac{k'(\lambda_2)}{k'(\lambda_{1,1})}\frac{v(\lambda_{2}) - v(\lambda_{1,2})}{v(\lambda_{1,1}) - v(\lambda_{1,2})}, \qquad \bar{\alpha}_{1,2} = - \alpha_{2} \frac{k'(\lambda_2)}{k'(\lambda_{1,2})}\frac{v(\lambda_{2}) - v(\lambda_{1,1})}{v(\lambda_{1,2}) - v(\lambda_{1,1})}.
\end{align}
We can now use the following property of $\delta$-functions
\begin{equation}
    \delta(f(x,y))\delta(g(x, y)) = \frac{\delta(x-\bar{x})\delta(y-\bar{y})}{| \det J|}, \qquad J = 
    \begin{pmatrix} 
        \partial_x f(x,y) & \partial_y f(x,y) \\
        \partial_x g(x,y) & \partial_y g(x,y)
    \end{pmatrix},
\end{equation}
where $(\bar{x}, \bar{y})$ is a zero of $(f(x,y), g(x,y))$. This gives for the $\delta$-functions implementing the conservation of energy-momentum the following expression
\begin{equation}
    \delta\big(k_{2 \rm ph}(\mathbf{p}_1,\mathbf{h}_1)+k_{1 \rm ph}(\mathbf{p}_2,\mathbf{h}_2)\big) \, \delta\big(\omega_{2 \rm ph}(\mathbf{p}_1,\mathbf{h}_1)+\omega_{1 \rm ph}(\mathbf{p}_2,\mathbf{h}_2)\big) = \frac{\delta(\alpha_{1,1} - \bar{\alpha}_{1,1})\delta(\alpha_{1,2} - \bar{\alpha}_{1,2})}{k'(\lambda_{1,1}) k'(\lambda_{1,2})|v(\lambda_{1,1}) - v(\lambda_{1,2})|}.
\end{equation}
After these preparational steps we evaluate the integrals over $\alpha_{1,1}$ and $\alpha_{1,2}$ with the help of the Dirac $\delta$-functions. This yields
\begin{equation}
\label{eqSM:mncontributionv_repeated}
\begin{aligned}
    \Gamma_{nm}=&\tau_0^{-1} \times \frac{(2 \pi)^2}{8} \int \dd \lambda_{1,1} \dd \lambda_{1,2} \dd \lambda_{2} \dd \alpha_2 \,
     \frac{G(\bfp_1, \bfh_1, \bfp_2, \bfh_2)}{k'(\lambda_{1,1}) k'(\lambda_{1,2})|v(\lambda_{1,1}) - v(\lambda_{1,2})|} \times \\
    &\Big( (h_n^{\rm Dr})'(\lambda_{1,1})\bar{\alpha}_{1,1}+(h_n^{\rm Dr})'(\lambda_{1,2})\bar{\alpha}_{1,2}+(h_n^{\rm Dr})'(\lambda_{2})\alpha_{2} \Big) \times \\
    &\Big(
    (h_m^{\rm Dr})'(\lambda_{1,1}) \bar{\alpha}_{1,1}+(h_m^{\rm Dr})'(\lambda_{1,2})\bar{\alpha}_{1,2}+(h_m^{\rm Dr})'(\lambda_{2})\alpha_{2} 
 \Big).
\end{aligned}
\end{equation}

The second step is to evaluate the integral over $\alpha_2$ which is proportional to the momentum transferred between the tubes. To this end we observe that the combinations of functions appearing in the last two lines of $\Gamma_{nm}$ can be rewritten according to
\begin{equation}
    \bar{\alpha}_{1,1}\varphi(\lambda_{1,1}) + \bar{\alpha}_{1,2} \varphi(\lambda_{1,2}) + \alpha_2 \varphi(\lambda_2) = \frac{\alpha_2 k'(\lambda_2)}{v(\lambda_{1,1}) - v(\lambda_{1,2})} \mathcal{S}[\varphi],
\end{equation}
where we introduced symbol $\mathcal{S}[\varphi]$ defined by
\begin{equation}
	\mathcal{S}[\varphi] = \frac{\varphi'(\lambda_{1,1})}{k'(\lambda_{1,1})}(v(\lambda_{1,2}) - v(\lambda_2)) + \frac{\varphi'(\lambda_{1,2})}{k'(\lambda_{1,2})}(v(\lambda_{2}) - v(\lambda_{1,1})) + \frac{\varphi'(\lambda_{2})}{k'(\lambda_{2})}(v(\lambda_{1,1}) - v(\lambda_{1,2})).
\end{equation}
We note that $\mathcal{S}[\varphi]$ is zero whenever any two of the $3$ variables $\lambda_{1,1}$, $\lambda_{1,2}$ or $\lambda_2$ are equal. Moreover, it also vanishes identically for constant $\varphi$ or $\varphi(\lambda)=k(\lambda), \omega (\lambda)$, and these functions naturally correspond to the collision invariants.
In the limit of small momentum transfer between the tubes we can also simplify function $G$. The density factors and the form-factors are independent of $\alpha_2$ in this limit. The density factors are $J_{m{\rm ph}}(\bfp, \bfh) = J_{m{\rm ph}}(\bfl, \bfl)$,
while the form-factors yield~\cite{DeN2015,DeN2018,Panfil2021}
\begin{align}
    F_{1 \rm ph}(\mathbf{p}_2,\mathbf{h}_2) &= k'(\lambda_2), \\
    F_{2 \rm ph}(\mathbf{p}_1,\mathbf{h}_1) &= \frac{2 \pi k^2 \mathcal{T}^{\rm dr}(\lambda_{1,1},\lambda_{1,2}) }{ k'(\lambda_{1,1})k'(\lambda_{1,2})\bar{\alpha}_{1,1}\bar{\alpha}_{1,2}} = - 2\pi \mathcal{T}^{\rm dr}(\lambda_{1,1},\lambda_{1,2}) \frac{(v(\lambda_{1,1}) - v(\lambda_{1,2}))^2}{(v(\lambda_2) - v(\lambda_{1,2}))(v(\lambda_2) - v(\lambda_{1,1}))},
\end{align}
with $k=k'(\lambda_{1,1})\bar{\alpha}_{1,1} + k'(\lambda_{1,2})\bar{\alpha}_{1,2}$. Here, $\mathcal{T}^{\rm dr}$ is the dressed integral kernel defined in \eqref{eqSM:Tdressed}. With this we find
\begin{equation}
\label{eqSM:mncontributionv_repeated}
\begin{aligned}
    \Gamma_{nm}=&\tau_0^{-1} \times \frac{(2 \pi)^4}{8} \int \dd \lambda_{1,1} \dd \lambda_{1,2} \dd \lambda_{2} \dd \alpha_2 \,
    \tilde{V}^2\big(k'(\lambda_2) \alpha_2\big) 
    \, J_{2 \rm ph}(\bfl_1, \bfl_1) \, J_{1 \rm ph}(\lambda_2, \lambda_2) \times \\
     & \frac{\alpha_2^2 k'(\lambda_2)^4 \mathcal{T}^{\rm dr}(\lambda_{1,1},\lambda_{1,2})^2}{k'(\lambda_{1,1}) k'(\lambda_{1,2})} \frac{|v(\lambda_{1,1}) - v(\lambda_{1,2})|}{(v(\lambda_2) - v(\lambda_{1,2}))^2(v(\lambda_2) - v(\lambda_{1,1}))^2} \mathcal{S}[h_n^{\rm Dr}]  \mathcal{S}[h_m^{\rm Dr}].
\end{aligned}
\end{equation}
The integral over $\alpha_2$ can be now performed. The result is the following
\begin{equation}
\label{eqSM:mncontributionv_repeated}
\begin{aligned}
    \Gamma_{nm}&=\tau_0^{-1} \times\frac{(2 \pi)^4 \tilde{V}^2}{8} \int \dd \lambda_{1,1} \dd \lambda_{1,2} \dd \lambda_{2} \frac{k'(\lambda_2) \mathcal{T}^{\rm dr}(\lambda_{1,1},\lambda_{1,2})^2 J_{2 \rm ph}(\bfl_1, \bfl_1) \, J_{1 \rm ph}(\lambda_2, \lambda_2))}{k'(\lambda_{1,1}) k'(\lambda_{1,2})} \times \\
      &\frac{|v(\lambda_{1,1}) - v(\lambda_{1,2})|}{(v(\lambda_2) - v(\lambda_{1,2}))^2(v(\lambda_2) - v(\lambda_{1,1}))^2} \mathcal{S}[h_n^{\rm Dr}]  \mathcal{S}[h_m^{\rm Dr}],
\end{aligned}
\end{equation}
where $ \tilde{V}^2 = \int {\rm d}k \tilde{V}^2\big(k\big) k^2$.
It is convenient to rewrite $\Gamma_{nm}$ in the following form, which appears in the main text
\begin{equation}
	\Gamma_{nm} =\frac{2 \pi}{\tau}  \int {\rm d}\bfl_1{\rm d}\bfl_2\, \mathcal{T}^{\rm dr}(\lambda_{1}, \lambda_{2})^2 |v(\lambda_{1}) - v(\lambda_{2})| g_{nm}(\lambda_1, \lambda_2),
\end{equation}
where $\tau^{-1}= \frac{\tilde{V}_{\rm inter}^2m}{4\hbar^3}$ and $\tilde{V}_{\rm inter}^2 = V_0^2 \tilde{V}^2$, moreover
\begin{equation}
	g_{nm}(\lambda_1, \lambda_2) =  \int {\rm d}\boldsymbol{\lambda}\, (k'(\lambda))^2 \frac{ \mathcal{S}[h_n^{\rm Dr}]  \mathcal{S}[h_m^{\rm Dr}]}{(v - v_{1})^2(v - v_{2})^2}.
\end{equation}
where the measure is $\dd \bfl= 2 \pi \rho_{\rm p}(\lambda) \rho_{\rm h}(\lambda)/k'(\lambda) \dd \lambda = \rho_{\rm p}(\lambda) f(\lambda) \dd \lambda$.

The apparent poles in $\Gamma_{nm}$ at $\lambda_{1,1} = \lambda$ and $\lambda_{1,2} = \lambda$ are regularized by the expressions in the brackets which vanish in those limits. The double limit when all the $\lambda$'s are equal is additionally regularized (and vanishes) because of the factor $|v(\lambda_{1,1}) - v(\lambda_{1,2})|$ in the numerator.

Explicitly, we check (we shorten the notation and introduce $\tilde{\varphi}=\varphi'/k'$, moreover $v = v(\lambda)$ and $v_i=v(\lambda_i)$, similarly for $\tilde{\varphi}$)
\begin{equation}
\begin{aligned}
	\lim_{v_{2} \rightarrow v} \frac{S[\varphi]}{v_{2} - v} &= \tilde{\varphi}_{1} + \lim_{v_{2} \rightarrow v} \frac{\tilde{\varphi}_{2} (v - v_{1}) + \tilde{\varphi} (v_{1} - v_{2})}{v_{2} - v} \\
	&= \tilde{\varphi}_{1} - v_{1} \lim_{v_2 \rightarrow v} \frac{\tilde{\varphi}_{2} - \tilde{\varphi}}{v_{2} - v} + \lim_{v_{2} \rightarrow v} \frac{\tilde{\varphi}_{2} v - \tilde{\varphi} v_{2}}{v_{2} - v} \\
	&= \tilde{\varphi}_{1} - \tilde{\varphi} - ( v_{1} - v)\tilde{\varphi}'.
\end{aligned}
\end{equation}
The other apparent pole at $v = v_1$ is regularized in the same way. This a consequence of symmetry of $\mathcal{S}$ under a cyclic transformation of variables, $\mathcal{S}[\varphi](\lambda, \lambda_1, \lambda_2) = \mathcal{S}[\varphi](\lambda_2, \lambda, \lambda_1) = \mathcal{S}[\varphi](\lambda_2, \lambda, \lambda_1)$. 

Finally, we also write down expression in large $c$ expansion, where the collision integral simplifies
\begin{equation}
\label{eqSM:mncontributionv_TG}
\begin{aligned}
    \Gamma_{nm}=&\frac{(2 \pi)^2}{\tau} \frac{4}{c^2} \int \dd \lambda_{1,1} \dd \lambda_{1,2} \dd \lambda_{2}  \rho_{\rm p}(\lambda_{1,1})\rho_{\rm h}(\lambda_{1,1})\rho_{\rm p}(\lambda_{1,2})\rho_{\rm h}(\lambda_{1,2})\rho_{\rm p}(\lambda_{2})\rho_{\rm h}(\lambda_{2})|\lambda_{1,1} - \lambda_{1,2}|\times \\
    &\Bigg( h_n'(\lambda_{1,1})\frac{1}{\lambda_2 - \lambda_{1,1}} + h_n'(\lambda_{1,2})\frac{1}{\lambda_{1,2}-\lambda_2}-h_n'(\lambda_{2})\left( \frac{1}{\lambda_2-\lambda_{1,1}}+ \frac{1}{\lambda_{1,2}-\lambda_2} \right) \Bigg) \times \\
    &\Bigg( h_m'(\lambda_{1,1})\frac{1}{\lambda_2 - \lambda_{1,1}} + h_m'(\lambda_{1,2})\frac{1}{\lambda_{1,2}-\lambda_2}-h_m'(\lambda_{2})\left( \frac{1}{\lambda_2-\lambda_{1,1}}+ \frac{1}{\lambda_{1,2}-\lambda_2} \right) \Bigg).
\end{aligned}
\end{equation}

\end{document}